\documentclass[a4paper,11pt]{article}
\usepackage{jcappub} 
\usepackage{lineno}
\usepackage{comment}

\newcommand{\mpch}{{\rm Mpc}~h^{-1}}

\newcommand{\be}{\begin{equation}}
\newcommand{\ee}{\end{equation}}

\newcommand{\bk}{\boldsymbol{k}}

\newcommand{\bfk}{\mbox{\boldmath$k$}}

\newcommand{\mksym}[1]{\ifmmode {\rm #1}\else #1\fi}

\newcommand{\leftparbox}[2]{\parbox{#1}{\begin{flushleft} #2 \end{flushleft}}}

\newcommand{\twoonesig}[4][\pbwidth]{
\begin{equation}
\left.
 \begin{aligned}
#2 \\ #3
 \end{aligned}
\ \right\} \ \ \mbox{\text{\leftparbox{#1}{#4}}}
\end{equation}
}

\title{\boldmath The simple way to measure evolving dark energy without prior-volume effects}

\author[a,b]{Maria Tsedrik,}
\author[c,a]{Pedro Carrilho,}
\author[d,e,f]{Chiara Moretti}

\affiliation[a]{Institute for Astronomy, The University of Edinburgh,
  Royal Observatory, Edinburgh EH9 3HJ, UK}

\affiliation[b]{Higgs Centre for Theoretical Physics, School of
  Physics and Astronomy, The University of Edinburgh, Edinburgh EH9
  3FD, UK}

\affiliation[c]{Centre for Astrophysics Research, University of Hertfordshire, College Lane, Hatfield AL10 9AB, UK}  

\affiliation[d]{INAF – Osservatorio Astronomico di Trieste, Via
  Tiepolo 11, I-34143 - Trieste, Italy}

\affiliation[e]{Institute for Fundamental Physics of the Universe, Via Beirut 2, 34151 Trieste, Italy}

\affiliation[f]{Istituto Nazionale di Fisica Nucleare, Sezione di Trieste,  via  Valerio  2,  34127 Trieste,  Italy}

\emailAdd{mtsedrik@ed.ac.uk,
  p.carrilho@herts.ac.uk, chiara.moretti@inaf.it}

\abstract{
 We present a simple yet effective method to resolve prior-volume effects, also known as projection effects, in full-shape analyses of the power
spectrum multipoles within the Effective Field Theory of Large-Scale Structure (EFTofLSS). By re-defining the EFTofLSS nuisance parameters to incorporate the contribution from the parameters impacting the amplitude of the EFTofLSS modelling components, we substantially mitigate projection effects. With the re-parametrisation the actual posterior maximum values are within the marginalised credible interval, eliminating significant shifts observed in the baseline analysis. We demonstrate the robustness of this method in full-shape $w_0w_a$CDM analyses on synthetic data in BOSS DR12 and DESI DR1 setups. For the evolving dark energy model, we then analyse the BOSS DR12 measurements, in combination with BAO information (from BOSS DR12, 6DF, SDSS DR7 MGS and eBOSS DR16 surveys) and 3$\times$2pt measurements from DES Y3 -- all data combinations are converging into the $w_0-w_a$ parameter region preferred by DESI+CMB+SNIa. From total combination of these large-scale structure probes without additional CMB information we find $w_0=-0.72 \pm 0.21, \, w_a=-0.91^{+0.78}_{-0.64}$. Despite the low significance of deviation from standard cosmology, this result underscores the potential of our re-parametrisation approach in delivering low-redshift cosmological constraints. We argue for the use of this approach in spectroscopic Stage IV surveys, where the potential deviation from standard cosmology can be detected with higher significance. }

\begin{document}
\maketitle
\flushbottom

\section{Introduction}
\label{sec:intro}

Spectroscopic surveys of Large-Scale Structure (LSS), such as the Baryon
Oscillation Spectroscopic Survey (BOSS) \cite{BOSS:2012dmf}, its extension eBOSS \cite{eboss2021}, and recently the Dark Energy Spectroscopic Instrument (DESI), \cite{desi2016} measured redshifts of millions of galaxies. These measurements can be translated into spatial distribution of galaxies, from which we can learn key cosmological information. From a ``bump", an excess of signal in the two-point correlation function of these galaxies measured at different redshifts, we can measure cosmological expansion and matter density \cite{Peebles1980, Eisenstein:1998tu}. Such excess of signal is known as baryonic acoustic oscillations (BAO), and can be used as a standard ruler: its position encodes the size of the Universe at the epoch slightly after re-combination. Beyond this feature, we can explore the shape of two-point correlators in redshift space and their Fourier-transform, the power spectrum multipoles. From these we can gain additional information on the growth of cosmic structures, the amplitude and shape of the primordial power spectrum \cite{Alam:2020jdv,Huterer:2022dds}. 

While in principle containing more cosmological information than BAO, a full-shape (FS) analysis also demands a more complex modelling prescription. Currently, the standard theoretical framework for modelling power spectrum multipoles in a FS analysis is the effective field theory of large-scale structure (EFTofLSS) \cite{baumann2012, carrasco2012,perko2016, delabella2017, DAmico:2019fhj, Nunes:2022bhn}. The EFTofLSS approach allows us to model the power spectrum multipoles in redshift space up to mildly nonlinear scales. Based on standard perturbation theory of structure formation, it includes contributions from smaller astrophysical scales via a small number of additional parameters, the so-called counterterms. In addition to the counterterms, the model includes other nuisance parameters such as those related to galaxy bias and shot noise contributions. Although these parameters allow the EFTofLSS to successfully fit a broad range of cosmologies (e.g., modified gravity \cite{Piga:2022mge,Taule:2024bot}, dark energy models \cite{DAmico:2020kxu,Gsponer:2023wpm}, various galaxy-halo connections \cite{Ivanov:2024hgq}), they also introduce projection or prior-volume effects, which complicate the interpretation of cosmological inferences. Projection effects arise when a high-dimensional non-Gaussian posterior distribution is compressed in a lower-dimensional parameter space \cite{Carrilho:2022mon, Simon:2022lde}. They are manifested by the best-fit, i.e. the Maximum \textit{A Posteriori} (MAP) value being in disagreement with the peak of the marginalised posterior \cite{gomezvalent2022,Hadzhiyska:2023wae}. This disagreement appears when unconstrained or weakly constrained regions of the nuisance parameter space significantly contribute to the marginalised posterior of cosmological parameters. Projection effects within EFTofLSS are especially prominent for beyond-$\Lambda$CDM cosmologies, in which additional parameters are strongly degenerate with cosmological and nuisance parameters \cite{Moretti:2023drg, maus2024}. Potential solutions to mitigate projection effects include various methods, such as informative priors on cosmological parameters motivated by CMB measurements \cite{Carrilho:2022mon, Moretti:2023drg}; Jeffreys priors \cite{Jeffreys:1946} on nuisance parameters \cite{Donald-McCann:2023kpx,Zhao:2023ebp,Paradiso:2024yqh}; HOD-informed priors on nuisance parameters \cite{Zhang:2024thl, DESI:2025wzd}; combination with external probes \cite{Tsedrik:2025cwc}.

In this work we propose a simple solution -- a re-parametrisation scheme\footnote{For a simpler toy-example of the re-parametrisation and its impact on prior-volume effects see the corresponding discussion in section 2.3 of ref.~\cite{Hadzhiyska:2023wae}.} for the nuisance parameters of the EFTofLSS. A similar approach is presented in the baseline analysis of DESI DR1 \cite{maus2024, DESI:2024jis} and photometric clustering with perturbative galaxy bias with DES Y3 \cite{DES:2021zxv, DES:2021bpo}. There, the nuisance parameters absorb $\sigma_8$, a parameter that corresponds to the normalisation of the power spectrum defined as the variance of the density field smoothed within a radius $R = 8$ $\mpch$. Such re-parametrisation helps connecting the model parameters with the observed power spectrum multipoles (or angular correlation functions in case of DES) and reducing biases from projection effects. Following the same logic, we propose that absorbing the Alcock-Paczynski (AP) \cite{alcock1979} amplitude significantly reduces projection effects for dark energy models beyond cosmological constant in FS analysis. The AP effect introduces anisotropies in the measured multipoles because of a discrepancy between the fiducial cosmology, assumed to convert redshifts to distances, and the true underlying cosmology. It also re-scales the overall amplitude of the signal.

This paper combines two topics: 1) mitigating projection effects via re-parametrisation of the EFTofLSS parameters and 2) constraining evolving dark energy with Stage III (pre-DESI) LSS data. The first topic is covered in sections~\ref{sec:projections} and \ref{sec:reparam}. There, we
discuss prior-volume effects in FS analysis (section~\ref{sec:projections}). We also study various re-parametrisations of the EFTofLSS nuisance parameters in a FS analysis with synthetic Stage III data (section~\ref{sec:fs-boss}) and then apply the best solution to a synthetic Stage IV scenario (section~\ref{subsec:stage4}). The second topic of this paper is covered in sections~\ref{sec:data-setup},  \ref{sec:fs-boss} and \ref{sec:external-probes}. We specify the different sets of Stage III data and the setup used in our analysis in section~\ref{sec:data-setup}. We provide constraints for the evolving dark energy parameters in BOSS DR12 FS analysis with our re-parametrisation in section~\ref{sec:fs-boss}. Then in sections~\ref{sec:boss-bao}-\ref{sec:extbao-des}, we add information from BAO (BOSS and external) and photometric probes (DES Y3) to gain more information from available LSS probes and provide constraints independent from the cosmic microwave background (CMB) and supernova (SN) measurements.

\section{Data and analysis setup}
\label{sec:data-setup}
\begin{figure}[t]
\centering
\includegraphics[width=\linewidth]{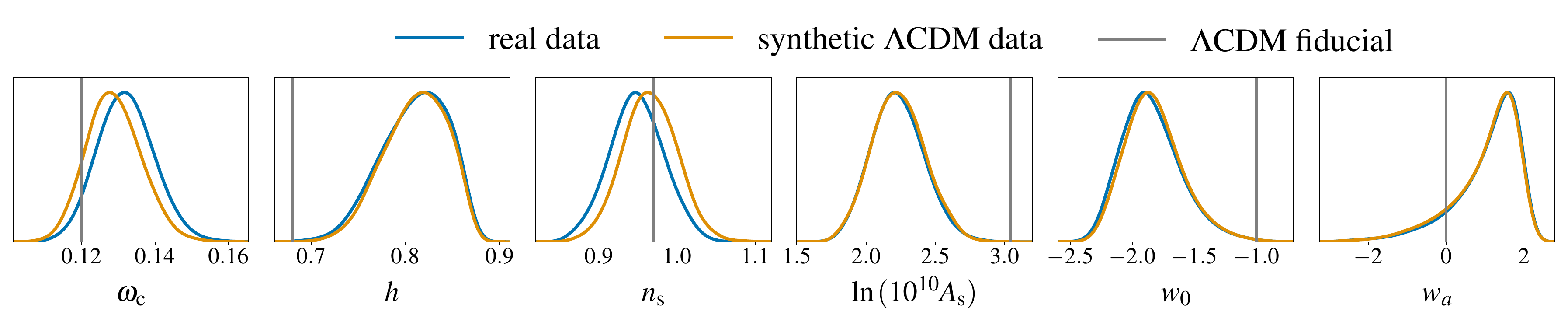}
\caption{Baseline FS analysis: 1D marginalised constraints on cosmological parameters on real BOSS DR12 data (blue) and synthetic $\Lambda$CDM data (orange). Fiducial values of cosmological parameters in the synthetic data are shown by the grey lines.}
\label{fig:synth-vs-real}
\end{figure}
We first summarise all the cosmological probes and analysis setups used in our inference. Note that, if not explicitly stated otherwise, we use the \texttt{Nautilus}-sampler \cite{Lange:2023} to explore the parameter space. Additionally, marginalised posteriors and credible intervals are obtained using \texttt{GetDist}\footnote{\url{https://github.com/cmbant/getdist}} \cite{getdist}, while for all MAP-values of the un-marginalised posterior distribution we use a minimiser called \texttt{minuit}\footnote{\url{https://github.com/jpivarski/pyminuit}}.\\
\textbf{BOSS:} The main focus of this work is FS analysis, for it we consider the galaxy power spectrum multipoles from BOSS DR12 \citep{alam2015, gilmarin2016, beutler2017}. 
The dataset consists of two galaxy samples: CMASS and LOWZ with effective redshifts $z_1=0.38$ and $z_3=0.61$, respectively. Each sample covers two different sky cuts, NGC in the north and SGC in the south. Hence in total, we fit 4 independent sets of multipoles. The multipole measurements and covariance\footnote{\url{https://github.com/oliverphilcox/full_shape_likelihoods/tree/main/data}} are provided in ref.~\cite{philcox2022} and are obtained with the windowless estimator \citep{philcox2021a,  philcox2021b}. The covariance is computed from ``MultiDark-Patchy" mock catalogues  \cite{kitaura2016, rodrigueztorres2016}. Complementarily, we also use four BAO measurements from BOSS DR12 in the same patches of sky \cite{philcox2022} (with the corresponding cross-covariance when combined with the multipoles). Further in the text, ``FS'' analysis includes only the multipoles, ``FS+BAO'' and ``BOSS'' denote the analysis with the multipoles in combination with the BAO data. In terms of the analysis choices, we keep priors on cosmological parameters broad and only impose Gaussian priors on the baryon density (BBN prior \cite{aver2015, cooke2018, schoneberg2019}) and spectral index (10 times the standard deviation from Planck's constraints \cite{planck2018cosmo}, similar to the DESI DR1 setup). Priors on the nuisance parameters in the baseline analysis are given in the left part of table~\ref{tab:param-priors}. Priors on the evolving dark energy parameters are uniform: $w_0 \in [-3, -0.3]$, $w_a \in [-3, 3]$, and with the following condition: $w_0+w_a<0$. We fit the power spectrum multipoles ($P_0, \,P_2, \,P_4$) with the identical scale-cuts of $k_{\rm max}=0.2~h/$Mpc. Our analysis setup and likelihood pipeline are discussed in detail in refs.~\cite{Carrilho:2022mon, Moretti:2023drg}.\\
\textbf{Synthetic data (Stage III):} To study projection effects in section~\ref{sec:fs-boss}, we create synthetic noiseless data vectors of BOSS-like power spectrum multipoles. In the synthetic data analysis we use the same covariance as in the BOSS DR12 real data analysis. For fiducial cosmological parameters we take
$\omega_{\rm c}=0.12, \, \omega_{\rm b}=0.02268, \, h=0.68, \, n_{\rm s}=0.97, \, \ln{(10^{10}A_{\rm s})}=3.044, \, w_0=-1, \, w_a=0, \, M_{\nu}=0.06~{\rm eV}$, while the
nuisance parameters are determined by maximising the full likelihood with
cosmology fixed to the fiducial values. In figure~\ref{fig:synth-vs-real} we show the agreement in the baseline FS analysis for evolving dark energy on real data (blue) and synthetic $\Lambda$CDM data (orange). 
As expected, we observe shifts towards lower $A_s$ values, which are present in the analyses of standard and extended cosmologies (see refs.~\cite{Carrilho:2022mon, Moretti:2023drg, Tsedrik:2025cwc}). We also observe shifts due to additional degeneracy between the evolving dark energy and expansion rate parameters: $h$ is shifted towards the upper prior bound, while the dark energy parameters prefer $w_0<-1$ and $w_a>0$. The driving mechanisms for both cases is discussed in the following section~\ref{sec:reparam}.\\
\textbf{Synthetic data (Stage IV):} To demonstrate robustness of our re-parametrisation approach in section~\ref{subsec:stage4}, we create synthetic noiseless data vectors of DESI DR1-like power spectrum multipoles. We follow the steps described in section 4.3. of ref.~\cite{Moretti:2023drg}: we assume the same fiducial values for the cosmological parameters as above and simulating the three galaxy
samples that are the target of DESI. This is repeated in two scenarios: standard cosmology ($w_0=-1, \,w_a=0$) and evolving dark energy ($w_0=-0.42, \,w_a=-1.75$). Values of effective redshifts and linear galaxy bias are taken from ref.~\cite{desi2016}, while number density of the samples and their total volume are computed with numbers from table~2 of ref.~\cite{DESI:2024aax}. 
\begin{table}[t]
  \centering
  \begin{tabular} { c  c  c  c  c}
    Galaxy sample & $z_{\rm eff}$ & $V_s~[{\rm Gpc}^3/h^3]$ &  $b_1$ & $\bar{n} \, [h^3/{\rm Mpc}^3]$\\
    \hline
    BGS   &  0.2  &  0.94 & 1.14 & $3.2 \times 10^{-4}$ \\
    LRGs  &  0.8  &  7.88  & 2.56  & $2.7 \times 10^{-4}$\\
    ELG   &  1.2  &  14.32  & 1.51  & $1.7 \times 10^{-4}$\\
    \hline
  \end{tabular}
  \caption{Parameters used to generate the synthetic DESI-like
    data-vectors in DR1 scenario. We follow ref.~\cite{desi2016} in
    computing the effective redshift $z_{\rm eff}$ and determining the linear bias $b_1$. Total volume and number density is computed with numbers from ref.~\cite{DESI:2024aax}.}
	\label{tab:desi-specs}
\end{table}
We list these numbers in
table~\ref{tab:desi-specs}. For the analysis we use the analytical covariance computed following ref.~\cite{taruya2010} and scale-cuts of $k_{\rm max}=0.25$ $h/{\rm Mpc}$ for all multipoles. \\
\textbf{extBAO:} In section~\ref{sec:external-probes}, we combine BOSS DR12 measurements with external BAO, namely with 
pre-reconstruction BAO measurements at low redshift from the 6DF survey \citep{beutler2011} 
and SDSS DR7 MGS \citep{ross2015}. 
We also add information from high redshift
measurements of the Hubble factor and angular diameter distance from
the Ly-$\alpha$ forest auto and cross-correlation with quasars
from eBOSS DR16 \citep{desbourboux2020}. \\
\textbf{DESY3:}
In section~\ref{sec:external-probes}, we also add 3$\times$2pt correlation functions from DES Y3 \cite{DES:2021wwk, DES:2022ccp}. They contain cosmic shear, galaxy clustering and galaxy-galaxy lensing information from sources in four redshift-bins and lenses from the first four redshift-bins of the MagLim sample \citep{DES:2020ajx}. The corresponding covariance matrix is obtained analytically as described and validated in ref.~\cite{DES:2020ypx}. We use the same scale-cuts as in the DES Y3 $\Lambda$CDM baseline analysis \cite{DES:2021rex} and model the nonlinear power spectrum with \texttt{HMcode2020} \cite{Mead:2020vgs}. We use the official DES-pipeline, \texttt{CosmoSIS}\footnote{\url{https://github.com/joezuntz/cosmosis-standard-library}}, in which we substitute the computation of linear and nonlinear matter power spectra with an emulator, \texttt{HMcode2020-emulator}\footnote{\url{https://github.com/MariaTsedrik/HMcode2020Emu}}, trained with \texttt{cosmopower}\footnote{\url{https://github.com/alessiospuriomancini/cosmopower}} \cite{SpurioMancini:2022}. We use the pipeline developed for the joint spectroscopic and photometric analysis from ref.~\cite{Tsedrik:2025cwc}. \\
\textbf{Planck+DESI+DESY5SN:} Finally, we repeat DESI DR2 BAO analysis \cite{DESI:2025zgx} in combination with the CMB (Planck 2018 low-$\ell$ TT and EE, CamSpec TT/TE/EE, and PR4 lensing \citep{Tristram:2023haj}) and DES Y5 SN \cite{DES:2024hip} using publicly available likelihoods in \texttt{Cobaya}\footnote{\url{https://github.com/CobayaSampler/cobaya}}. The constraints we obtain are in good agreement with the official DESI results from ref.~\cite{DESI:2025zgx}. The purpose of repeating this analysis is to plot the corresponding posteriors for the evolving dark energy parameters together with the results of our Stage III LSS-only analysis in fig.~\ref{fig:w0wa}.

\section{Prior-volume effects in full-shape analysis}
\label{sec:projections}

In our modelling of the power spectrum multipoles we follow a prescription analogous
to the one of ref.~\cite{chudaykin2020}, but employ the independent
implementation in the \texttt{PBJ} code. The latter has been extensively validated on N-body simulations in $\Lambda$CDM ~\cite{oddo2020, oddo2021, rizzo2023, Carrilho:2021hly, Tsedrik:2022cri}. Below we outline the key components relevant for the discussion on projection effects, but refer to refs.~\cite{Carrilho:2022mon, Moretti:2023drg} for
a more detailed description. The one-loop EFTofLSS galaxy power spectrum at a fixed redshift can be written as
\begin{flalign}
\label{eq:Pgg}
    P_{\rm gg}(k, \mu) =P_\mathrm{SPT}(\bk) + P_\mathrm{ctr}(\bk) + P_\mathrm{stoch} (\bk) \, .
\end{flalign}
The first term is a standard perturbation theory one-loop power spectrum
with the redshift space kernels from ref.~\cite{scoccimarro1999} and integrals of the linear power spectrum, $P_{\rm L}$, with various powers and combinations of galaxy bias parameters -- $b_1, \,b_2, \, b_{\mathcal  G_2}, \, b_{\Gamma_3}$ \cite{mcdonald2009,desjacques2018}. The term important for our discussion is
\begin{align}
P_{b_{\Gamma_3}} (k, \mu) = & b_{\Gamma_3} (b_1 + f \mu^2) P_{13}(k, \mu)\,,
\end{align}
where $P_{13}$ is a loop-correction integral proportional to $P_{\rm L}^2$, $\mu$ is the cosine of the angle between the wavevector $\bfk$
and the line of sight, $f$ is the growth rate.
The second term from eq.~\ref{eq:Pgg} is the EFTofLSS counterterm contribution:
\begin{align}
P_{\rm ctr}(k, \mu) = &- 2 \tilde{c}_0 k^2 P_{\rm L}(k) - 2 \tilde{c}_2 k^2 f
\mu^2 P_{\rm L}(k) - 2 \tilde{c}_4 k^2 f^2 \mu^4 P_{\rm L}(k)
\nonumber\\ &+c_{\nabla^4 \delta} k^4 f^4 \mu^4 (b_1 + f \mu^2)^2
P_{\rm L}(k)\,.
\label{eq:couterterms}
\end{align}
Following ref.~\cite{chudaykin2020}, we re-define the $k^2$-counterterm parameters ($\tilde{c}_0, \, \tilde{c}_2, \, \tilde{c}_4$) in order to have separate
contributions to each multipole.  Lastly, the third term from eq.~\ref{eq:Pgg} contains the shot-noise contribution:
\be P_{\rm stoch}(k, \mu) = N + e_0 k^2 + e_2 k^2\mu^2
\, ,
\label{eq:noise}
\ee
where $N$ is a constant that includes deviations from pure Poisson
shot noise, and we have two additional scale-dependent terms with $e_0$ and $e_2$ in units of $(\mpch)^2$. We include the impact of the fiducial cosmology assumed when
converting redshifts to distances in the data by correcting $(k, \mu)$ with AP distortions and re-scale $P_{\rm gg}$ with the AP amplitude \cite{alcock1979}. The AP amplitude, $A_{\rm AP}$ is a function of redshift and background cosmology:
\be A_{\rm AP}(z)=\left(\frac{H_0^{\rm
    fid}}{H_{0}}\right)^3\frac{H(z)}{H^{\rm fid}(z)}\left(\frac{D_{A}^{\rm
    fid}(z)}{D_A(z)}\right)^2\,  \ee
with $H_0$ being the Hubble factor today, $D_A$ the angular diameter
distance, and the superscript $^{\rm fid}$ referring to quantities evaluated in the
fiducial cosmology. Finally, we project the anisotropic power spectrum
to multipoles, $P_l(k)$, with Legendre polynomials.

To accelerate the convergence of MCMC chains, we perform analytical marginalisation of nuisance parameters that appear linearly in the model (counterterms, shot-noise and the third order bias parameter) \cite{damico2020, damico2021}. We  separate the power spectrum multipoles into analytically non-marginalised (nm) and marginalised (m) parts: 
\begin{align}
    P_l(k) &= P_l^{\rm nm}(k) + \sum_i n^{\rm m}_i P^{\rm m}_{i, l}(k)\, ,
\end{align}
where the nuisance parameters with superscript ``${\rm m}$'' appear linearly in the model and have Gaussian priors. For a Gaussian likelihood this allows us to solve the posterior integral over $n^{\rm m}_i$ analytically. The resulting marginalised log-posterior is a function of cosmological and non-marginalised nuisance parameters $(\Omega, n^{\rm nm})$ and proportional to
\begin{equation}
\label{eq:LaplaceApprox}
    \chi^2_{\rm m}(\Omega, n^{\rm nm}) = \chi^2_*(\Omega, n^{\rm nm})+\ln{\det{F_2}}(\Omega, n^{\rm nm})+\mathrm{const}\,.
\end{equation}
The first term on the r.h.s. is what we call the ``profile likelihood'', while the second term is called the Laplace term (see ref.~\cite{Hadzhiyska:2023wae} for details). 
They are given by
\begin{align}
    \chi^2_*&=F_0-F_{1, i} F^{-1}_{2, ij} F_{1, j}\, ,\\
        F_0 &= \Delta_l \mathrm{Cov}^{-1}_{ll'} \Delta_{l'}+\mu_i \mathrm{C}^{-1}_{ij} \mu_j\, ,\\
    F_{1, i}&= -P^{\rm m}_{i,l} \mathrm{Cov}^{-1}_{ll'} \Delta_{l'}+ \mathrm{C}^{-1}_{ij} \mu_j\, ,\\
    \label{eq:F2ij}
    F_{2, ij}&=P^{\rm m}_{i, l} \mathrm{Cov}^{-1}_{ll'}P^{\rm m}_{j, l'}+\mathrm{C}^{-1}_{ij}\,  
\end{align}
with $\Delta_l(k) = P_l^{\rm nm}(k) - D_l(k)$ being the difference of the non-marginalised part of the theoretical prediction with respect to the data, $\mathrm{Cov}$ being the covariance matrix for the data, and $\mathrm{C}_{ij}$ being the covariance for any possible Gaussian priors on nuisance parameters, with means $\mu_i$. For flat priors on nuisance parameters $\mathrm{C}^{-1}_{ij}=0$. The Laplace term is responsible for the prior-volume effects: the volume in the nuisance parameter space when integrated over fixed $(\Omega, n^{\rm nm})$. In other words, this term corresponds to the shift in the maximum of the marginalised distribution with respect to the maximum of the full distribution, because the process of marginalisation favours regions of parameter space that cover a larger volume of the probability density in the direction of integration. Improving the measurements (i.e. shrinking the covariance), or imposing informative priors on parameters makes the profile likelihood term larger and less sensitive to shifts due to the Laplace term. Note that applying Jeffreys priors on analytically marginalised parameters results in the cancellation of the Laplace term. In that case, instead of the Gaussian priors and $\mathrm{C}_{ij}$ in the analytically marginalised nuisance parameters, one applies Jeffreys priors $\propto \sqrt{|F(n_{\rm m})|}$ dependent on the Fisher information matrix, $F$. The Fisher matrix for these parameters is equivalent to the first term on the r.h.s. of eq.~\ref{eq:F2ij}, hence leading to the total cancellation of the Laplace term in the marginalised posterior distribution. In that scenario, prior-volume effects due to the parameters appearing linearly in the model are fully mitigated.
\begin{figure}[t]
\centering
\includegraphics[width=0.48\textwidth]{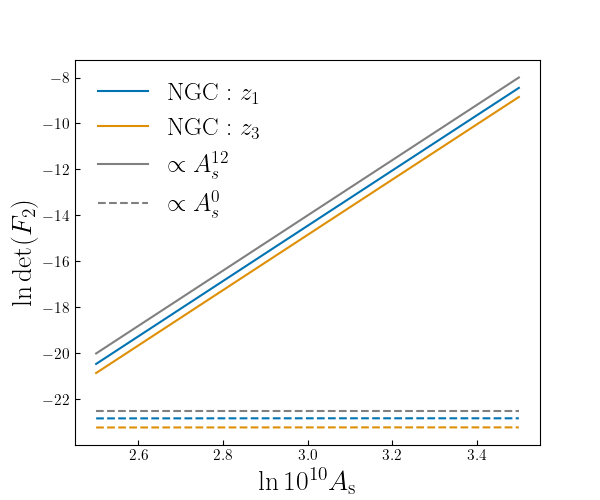}
\includegraphics[width=0.48\textwidth]{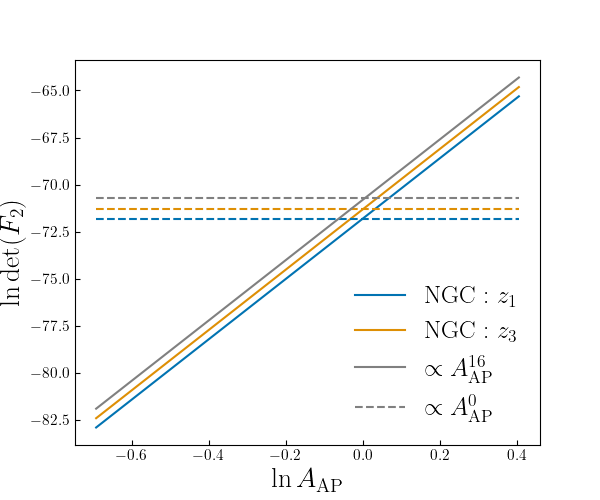}
\caption{Illustration of the Laplace term's dependence on two amplitude-controlling parameters, primordial amplitude $A_{\rm s}$ and AP amplitude $A_{\rm AP}$. The computation is done with the synthetic data for BOSS northern sky cuts (NGC) at two redshifts, $z_1=0.38$ and $z_3=0.61$. Cosmological and non-analytically marginalised nuisance parameters, $(\Omega, n^{\rm nm})$, are fixed to their fiducial values. Left: only the Fisher part of $F_2$ without noise-contribution. Dashed lines after the re-parametrisation $b_{\Gamma_3} \rightarrow A_s^2b_{\Gamma_3}$ and $c_i \rightarrow A_s c_i$ with $i \in \{0,2,4, \nabla^4\delta \}$. Right: only the Fisher part of $F_2$ with noise-terms. Dashed lines after the re-parametrisation $n_i \rightarrow A_{\rm AP} n_i$ with $n_i \in \{b_{\Gamma_3}, c_0,c_2,c_4, c_{\nabla^4\delta}, N, e_0, e_2 \}$.}
\label{fig:Laplace-amp}
\end{figure}

Many of the nuisance parameters of the EFTofLSS model are degenerate with other amplitudes that depend on cosmological parameters. A clear case is that of counterterms in eq.~\ref{eq:couterterms}, $P_{\rm ctr}\sim ck^2P_{\rm L}$, with $P_{\rm L}$ scaling like $A_{\rm s}$ (or $\sigma^2_8$). In general, this can be written as $P_{\rm ctr}=\alpha f(k)$ where $f(k)$ controls only the scale dependence of the counterterm, while $\alpha$ accounts for its overall amplitude and in the current example can be written as $\alpha = c A_{\rm s}$. It is clear that the model and the corresponding likelihood can only depend on the combination $\alpha$ and therefore that is the only parameter that gets directly constrained, while $c$ can only be measured with knowledge of the amplitude, $A_{\rm s}$. Providing a particular prior on $c$ can therefore provide information that is not obvious from a first glance. To see this, let us look at the infinitesimal posterior probability in the case where only $c$ and $A_{\rm s}$ are varied:
\begin{equation}
\mathrm{d}P = \mathrm{d}c\,\mathrm{d}A_{\rm s} \mathcal{L}(A_{\rm s},c)\,,
\end{equation}
where we assume a flat prior on both parameters. Instead of marginalising over $c$, let us first change variables from $c$ to $\alpha$:
\begin{equation}
\mathrm{d}P = \mathrm{d}\alpha\,\mathrm{d}A_{\rm s} \frac{1}{A_{\rm s}}\mathcal{L}(A_{\rm s},\alpha) = \mathrm{d}\Delta\alpha\,\mathrm{d}A_{\rm s} \frac{1}{A_{\rm s}}\mathcal{L}_1(A_{\rm s})\mathcal{L}_2(\Delta\alpha)\,.
\label{eq:dP_As}
\end{equation}
In this form, it is possible to separate the dependence on the amplitude from the dependence on $\Delta\alpha=\alpha-\alpha_*$, the difference of $\alpha$ with respect to its best-fit value. Marginalisation over $\Delta\alpha$ is then straightforward and will not affect the distribution of $A_{\rm s}$ further. This is the only form which allows us to make this separation and clearly identify the contributions from the likelihood, while extracting the effective prior on $A_{\rm s}$ that has been imposed by the original choice of a flat prior in terms of $c$. We see now more transparently that this flat prior is informative giving a prior in the form $1/A_{\rm s}$ (eq.~\ref{eq:dP_As}). This provides a preference for low values of the amplitude with which the nuisance parameter is degenerate, generating prior volume effects. It is also clear from this construction that the effect adds up with every additional free nuisance parameter that is degenerate with the same amplitude, giving rise to the large powers of $A_{\rm s}$ in the prior, which correspond to those in the Laplace term, as shown in the left panel of figure~\ref{fig:Laplace-amp} (solid lines). 

This argument applies more generally, affecting any other parameter that enters an amplitude that is degenerate with a nuisance parameter in this way. Another example is the amplitude factor due to the AP effect, which multiplies counter-terms and stochastic parameters in the same way. As before, setting flat priors for those nuisance parameters introduces additional information on this AP amplitude that causes the posterior to prefer lower values of $A_{\rm AP}$ (see the right panel of figure~\ref{fig:Laplace-amp}) and leading to further informative priors on the parameters on which it depends: $\Omega_{\rm m}$ or $\omega_{\rm c}-\omega_{\rm b}-h$ and the equation of state for dark energy $w(a)$. 

In the examples above, we identified amplitudes ($A_{\rm s}$, $\sigma_8^2$  and $A_{\rm AP}$) which impact the terms in question (e.g., the counter-terms) at all scales. In this case, eliminating projection effects in those amplitudes via re-parametrisation or via Jeffreys priors is identical: both approaches make the difference between the marginalised posterior and the profile-likelihood term in eq.~\ref{eq:LaplaceApprox} amplitude-independent. For other parameters, such as $\Omega_{\rm m}$ (see figure~6 in ref.~\cite{Hall:2021qjk}), their scale-dependent impact on the power spectrum implies that determining the best amplitude to re-parametrise becomes more complex. The analogous to $\alpha = c A_{\rm s}$ from the paragraph above, would replace $A_{\rm s}$ with the best-measured amplitude at the scales measured by the particular experiment. We argue below for motivated combinations of $\sigma_8$  and $A_{\rm AP}$ that correspond to this for each parameter in question, but a more precise method would be to use information on uncertainties of the measurements at different scales to determine this amplitude. This information, i.e. the covariance matrix of the data, is by construction included in Jeffreys priors via the Fisher matrix. Therefore, the Jeffreys priors approach is fundamentally a more nuanced yet experiment-dependent way to eliminate projection effects, being equivalent to some more complex re-parametrisation. The re-parametrisation of the amplitude-controlling parameters that we introduce here can be seen as its sub-class or a first-order approximation that is independent of the specific experiment, since the amplitudes we use are all found directly from the modelling and do not require covariance information.

\begin{figure}[t]
\centering
\includegraphics[width=\linewidth]{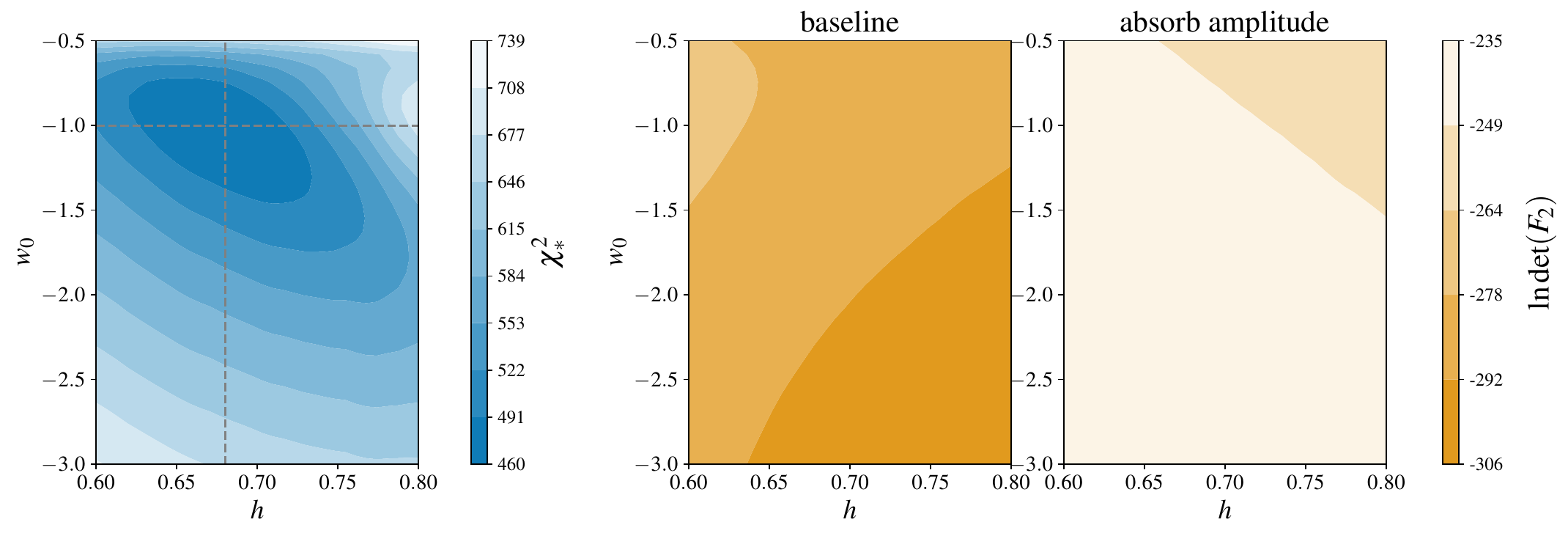}
\caption{Impact of the re-parametrisation on the Laplace term. The computation is done with the synthetic data for the full set of BOSS multipoles. Cosmological and non-analytically marginalised nuisance parameters, $(\Omega, n^{\rm nm})$, are fixed to their fiducial values. }
\label{fig:grids}
\end{figure}

In beyond-$\Lambda$CDM cosmologies, extended parameters impact the evolution of structure growth and expansion history, hence affecting the amplitude signal too. For extended cosmologies which only modify the growth of structure (e.g., modified gravity models with $\Lambda$CDM background), imposing an informative prior on the primordial amplitude (e.g., from CMB measurements) might be sufficient to resolve projection effects due to the degeneracy between the extended parameters and $A_{\rm s}$ \cite{Moretti:2023drg}. On the other hand, dark energy models modify the background expansion, opening an additional degeneracy direction and affecting the AP effect. In this scenario the arising projection effects cannot be resolved by fixing the primordial amplitude or re-parametrising nuisance parameters with $A_{\rm s}$ (or $\sigma^2_8$). While these projections do become slightly weaker with additional BAO information, they are not removed to a satisfying degree yet (see figure~3 of ref.~\cite{DESI:2024jis}). However, when we re-define nuisance parameters to incorporate all contributions to the amplitude signal, including the AP-amplitude, we reduce the dependence of the Laplace term on the degenerate parameters (see dashed lines in figure~\ref{fig:Laplace-amp}), and hence the corresponding projection effects. As argued above, flat priors on the analytically marginalised parameters introduce informative priors on the cosmological parameters contributing to the AP effect. Let us demonstrate exactly that in figure~\ref{fig:grids}. As we showed in the baseline analysis in figure~\ref{fig:synth-vs-real}, variation of the evolving dark energy parameters on noiseless $\Lambda$CDM data leads to strong biases in the expansion rate and dark energy parameters. When fixing all cosmological and non-analytically marginalised parameters to their fiducial values and varying only $w_0-h$, the ``profile likelihood'' term (or the log-posterior times minus two) evaluated at the best-fit values of the analytically marginalised parameters (left panel) clearly dips around the fiducial values. Nevertheless, the minimum of the overall sum from eq.~\ref{eq:LaplaceApprox} (i.e. the maximum of the marginalised posterior) is strongly affected by the Laplace term (middle panel), which drags it to lower values of $w_0$ and higher values of $h$. This dependency is weakened when the amplitude-impacting parameters get absorbed in the nuisance parameters that appear linearly in the modelling (right panel). As per discussion above, in contrast to Jeffreys priors, this does not make the Laplace term completely independent from the cosmological parameters in question. Nevertheless, it still significantly decreases the sensitivity to them, which supports our statement of the re-parametrisation being a first-order approximation of the Jeffreys priors approach. 

Additionally, fig.~\ref{fig:grids} can be useful to visualise the impact of projection effects on the width of the marginalised posteriors. If the contributions to the marginalised posterior distribution coming from the profile-likelihood term and from the prior-volume term have comparable significance and are in tension (i.e. they prefer different parts of the parameter space), then their combination might result not only in shifts of the parameter estimates, but also in tighter widths. In other words, the total marginalised posterior distribution has smaller volume than in the Jeffreys priors or re-parametrisation approaches. This does not mean that any information is lost when the re-parametrisation or Jeffreys priors are applied. On the contrary, broader constraints in this case would indicate that no additional information is imposed on the cosmological parameters by our choices in the nuisance parameters.

\section{Re-parametrisation in full-shape analysis}
\label{sec:reparam}
\subsection{BOSS DR12}
\label{sec:fs-boss}

\begin{table}[h!]
\footnotesize
  \centering
  \begin{tabular} { c  c | c c }
  \multicolumn{2}{c|}{{\bf baseline}} & \multicolumn{2}{c}{{\bf re-parametrisation}} \\ 
    parameter &  prior & parameter  &  prior \\
    \hline
	\hline
    $b_1$ & [0, 4] &  $b_1$ & [0, 4] \\
    $b_2$ & $\mathcal{N}$(0, 1) &  $b_2$ & $\mathcal{N}$(0, 1)  \\
    $b_{\mathcal{G}_2}$ & $\mathcal{N}$(0, 1) &  $b_{\mathcal{G}_2} $ & $\mathcal{N}$(0, 1) \\
    \hline
      $b_{\Gamma_3}$ & $\mathcal{N}\left(0, 1\right)$ &  $b_{\Gamma_3} A_{\rm AP} \times \{1, \tilde{A}^{2}_\mathrm{s}, \tilde{A}^{2}_{\mathrm{s}, z}, \sigma_{8, z}^4 \}$ & $\mathcal{N}\left(0,\{1, 3, 2, 1\}\right)$ \\
      $c_0$ & $\mathcal{N}(0, 30)$ &  $c_0 A_{\rm AP} \times \{1, \tilde{A}_\mathrm{s}, \tilde{A}_{\mathrm{s}, z}, \sigma_{8, z}^2 \}$ & $\mathcal{N}(0, \{30, 90, 60, 30\})$ \\
    $c_2$ & $\mathcal{N}(30, 30)$ & $c_2  A_{\rm AP}\times \{1, \tilde{A}_\mathrm{s}, \tilde{A}_{\mathrm{s}, z}, \sigma_{8, z}^2 \}$ & $\mathcal{N}(\{30, 60, 37.5, 15\}, \{30, 90, 60, 30\})$ \\
    $c_4$ & $\mathcal{N}(0, 30)$ &  $c_4 A_{\rm AP} \times \{1, \tilde{A}_\mathrm{s}, \tilde{A}_{\mathrm{s}, z}, \sigma_{8, z}^2 \}$ & $\mathcal{N}(0, \{30, 90, 60,  30\})$ \\
    $c_{\nabla^4\delta}$ & $\mathcal{N}(-10^3, 10^3)$ &  $c_{\nabla^4\delta} A_{\rm AP}\times \{1, \tilde{A}_\mathrm{s}, \tilde{A}_{\mathrm{s}, z}, \sigma_{8, z}^2 \}$ & $\mathcal{N}(-10^3 \cdot  \{1, 2, 1.25, 0.5\}, \{1, 3, 2, 1\}\cdot 10^3)$ \\
    $N$ & $\mathcal{N}\left(\frac{1}{\bar{n}},\frac{2}{\bar{n}}\right)$&$NA_{\rm AP}$ & $\mathcal{N}\left(\frac{1}{\bar{n}},\frac{2}{\bar{n}}\right)$\\
    $e_0$ & $\mathcal{N}\left(0,\frac{2}{\bar{n}k_{\rm
        NL}^2}\right)$  &$e_0 A_{\rm AP}$ & $\mathcal{N}\left(0,\frac{2}{\bar{n}k_{\rm
        NL}^2}\right)$ \\
    $e_2$ & $\mathcal{N}\left(0,\frac{2}{\bar{n}k_{\rm
    NL}^2}\right)$ &$e_2 A_{\rm AP}$ & $\mathcal{N}\left(0,\frac{2}{\bar{n}k_{\rm
    NL}^2}\right)$ 
  \end{tabular}

  \caption{Setup for runs from figure~\ref{fig:boss-anmarg-parameterisation}, where $\tilde{A}_\mathrm{s}= A_s \times 10^9$, $\tilde{A}_{\mathrm{s}, z}=D^2(z)/D^2_{\Lambda \rm CDM}(0) \times A_s \times 10^9$, $\sigma_{8,z}=D(z)/D(0) \sigma_8$. The horizontal line separates analytically marginalised parameters. Normal distributions follow linear transformation: for $X_{\rm baseline}\sim\mathcal{N}(\mu, \sigma)$, if $Y_{\rm re-parametrisation}=aX_{\rm baseline}$ then $Y_{\rm re-parametrisation}\sim\mathcal{N}(a\mu, a\sigma)$. We shift the means of the nuisance parameter priors with the rescaled fiducial values of the absorbed amplitude parameters: $A_{\rm AP, fid}=1$, $\tilde{A}_\mathrm{s, fid}\approx 2.1$, $\tilde{A}_{\mathrm{s}, z, \mathrm{fid}} \approx 1.25$, $\sigma^2_{8,z, \mathrm{fid}}\approx0.4$. We increase the variance of the new normal distributions by a factor of 2 (for re-scaling with $\tilde{A}_{\mathrm{s}, z}$) and 3 (for re-scaling with $\tilde{A}_{\mathrm{s}}$) to take into account values of the primordial amplitude allowed by the prior ranges in cosmological parameters.}
  \label{tab:param-priors}
\end{table}
\begin{figure}[h!]
\centering
\includegraphics[width=0.6\textwidth]{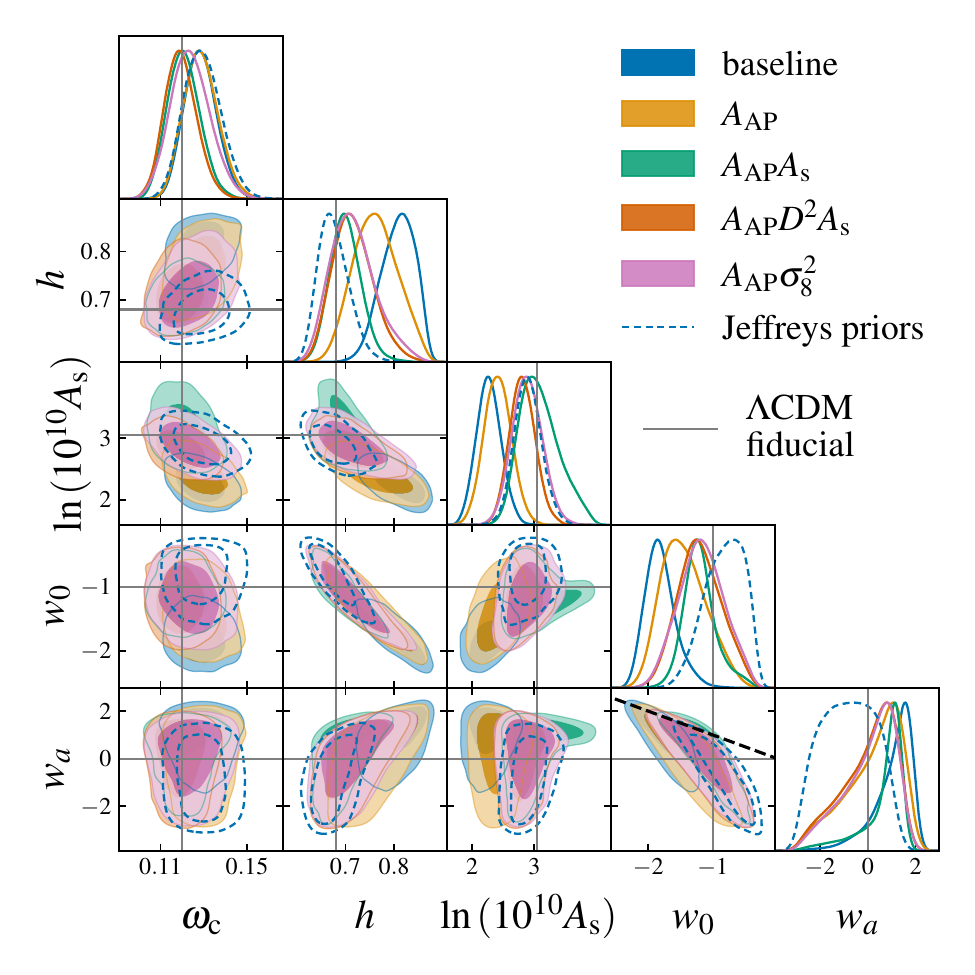}
\includegraphics[width=0.34\textwidth]{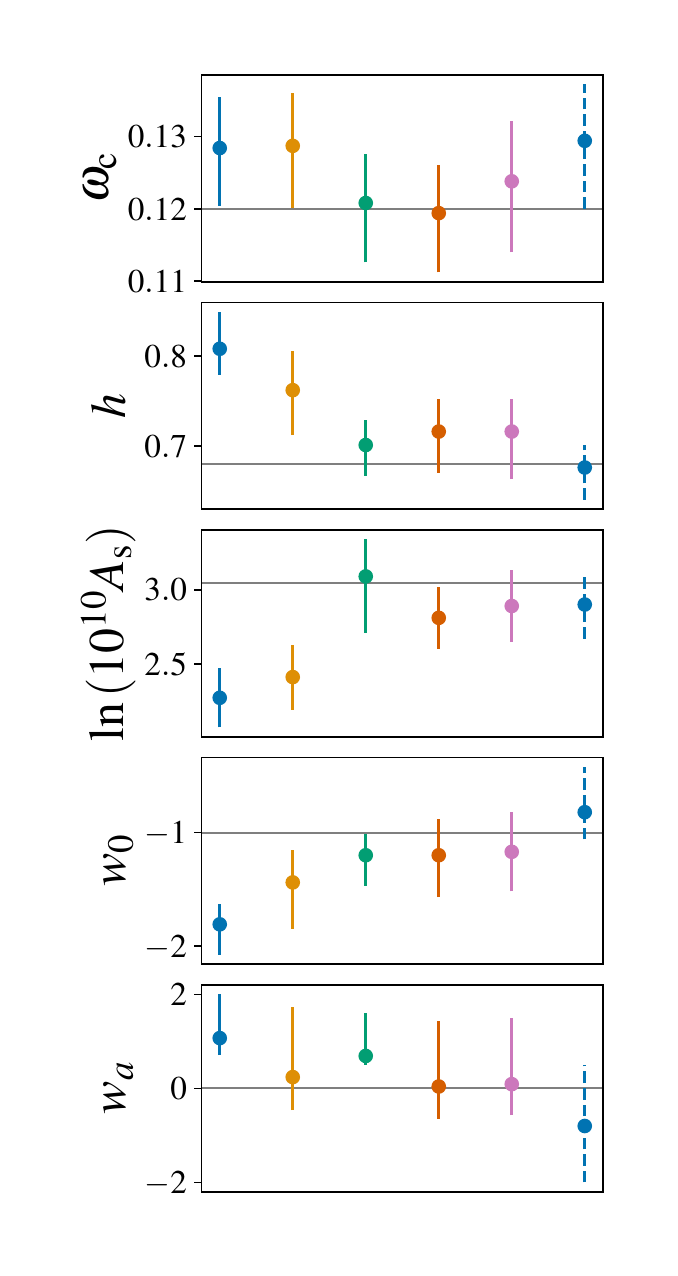}
\caption{Different re-parametrisations of the EFTofLSS nuisance parameters and their impact on cosmological constraints demonstrated on synthetic data. \textit{Left panel:} marginalised posterior distribution for the cosmological
      parameters in the $w_0w_a$CDM cosmology and the five re-parametrisation choices,
      as detailed in the legend (see table~\ref{tab:param-priors}). Grey solid lines mark the
      fiducial values of the noiseless synthetic data from a fit on the BOSS DR12 multipoles in a $\Lambda$CDM scenario. \textit{Right panel:} errorbars correspond to 68\% c.l. of 1D marginalised constraints on cosmological parameters from the left panel.}
\label{fig:boss-anmarg-parameterisation}
\end{figure}

After identifying and discussing the problem of projection effects in the previous section, we now investigate different amplitude controlling parameters. In this section we study how re-defining the nuisance parameters to take their impact into account mitigates the shifts in the maxima of un-marginalised and marginalised posterior distributions. The amplitude controlling parameters in question are the primordial amplitude $A_{\rm s}$, the AP amplitude $A_{\rm AP}$ (dependent on the background modifying parameters), the growth factor $D(z)$ (dependent on the structure growth modifying parameters), and $\sigma_8$ (a nonlinear function of all cosmological parameters). In table~\ref{tab:param-priors} we summarise different options of re-defining or re-parametrising nuisance parameters with the aforementioned amplitudes and the corresponding prior choices. In figure~\ref{fig:boss-anmarg-parameterisation} we investigate the impact of the re-parametrisation scenarios of the analytically marginalised nuisance parameters on the noiseless BOSS DR12 synthetic data, for which the fiducial underlying cosmology is known. First, we see that absorbing the AP amplitude alone (in yellow) already brings the degenerate trio of $h-w_0-w_a$ closer to their fiducial values. Second, addition of the primordial amplitude parameter (in green) improves the degeneracy $A_{\rm s}-h-\omega_{\rm c}$. Finally, addition of a redshift dependent components to the power spectrum amplitude, i.e. multiplication with the growth factor squared of $A_{\rm s}$ and $\sigma_8^2$ (in orange and pink, respectively) broadens the $h-w_0-w_a$ constraints. For the rest of our analysis in this paper we choose the re-parametrisation of the analytically marginalised parameters with the time-dependent $\sigma_8^2$ and the AP amplitude (pink contours). We base this decision not on the better performance of this particular re-parametrisation in the synthetic data tests, but on the fact that it  captures the amplitude signal at varying redshifts $P_\ell\propto A_{\rm AP}(z)\sigma^2(z) f(z,k)$ in a more physically motivated way than $P_\ell\propto A_{\rm AP}(z)D^2(z) A_{\rm s} f(z,k)$. Additionally, this re-parametrisation also proves to be more stable to the increasing prior volume due to slightly changing the degeneracy in the $A_{\rm s}-h-\omega_{\rm c}$ direction, since $\sigma_8$ is a nonlinear function of these parameters. For more details on prior-dependence see appendix~\ref{app:priors}.

In figure~\ref{fig:boss-anmarg-parameterisation} we also show the baseline analysis with the Laplace term removed via partial Jeffreys priors (dashed blue lines). Parameters $h$ and $A_{\rm s}$, as well as $w_0$ and $w_a$ recover the fiducial values correctly, without the previously observed biases in the baseline analysis (solid blue lines). There is a $\sim1\sigma$ bias in $\omega_{\rm c}$ with respect to the fiducial value. The fact that it is not present with the re-parametrisation is possibly coincidental. As argued in section~\ref{sec:projections}, with the re-parametrisation we do not make the Laplace term completely cosmology-independent, hence the residual projection effects from the analytically marginalised nuisance parameters can impact or even cancel out other intrinsic non-Gaussianities of the posterior. Overall, the fiducial cosmological parameters are always recovered within $1\sigma$ with Jeffreys priors and our chosen re-parametrisation approach. In both approaches the constraints on the evolving dark energy parameters are wider than in the baseline analysis. This demonstrates the additional consequence of projection effects discussed in the last paragraph of section~\ref{sec:projections} -- constraints typically appear narrower when strong projection effects exist, but are broader when spurious information is removed. 

\begin{figure}[h!]
\centering
\includegraphics[width=0.6\textwidth]{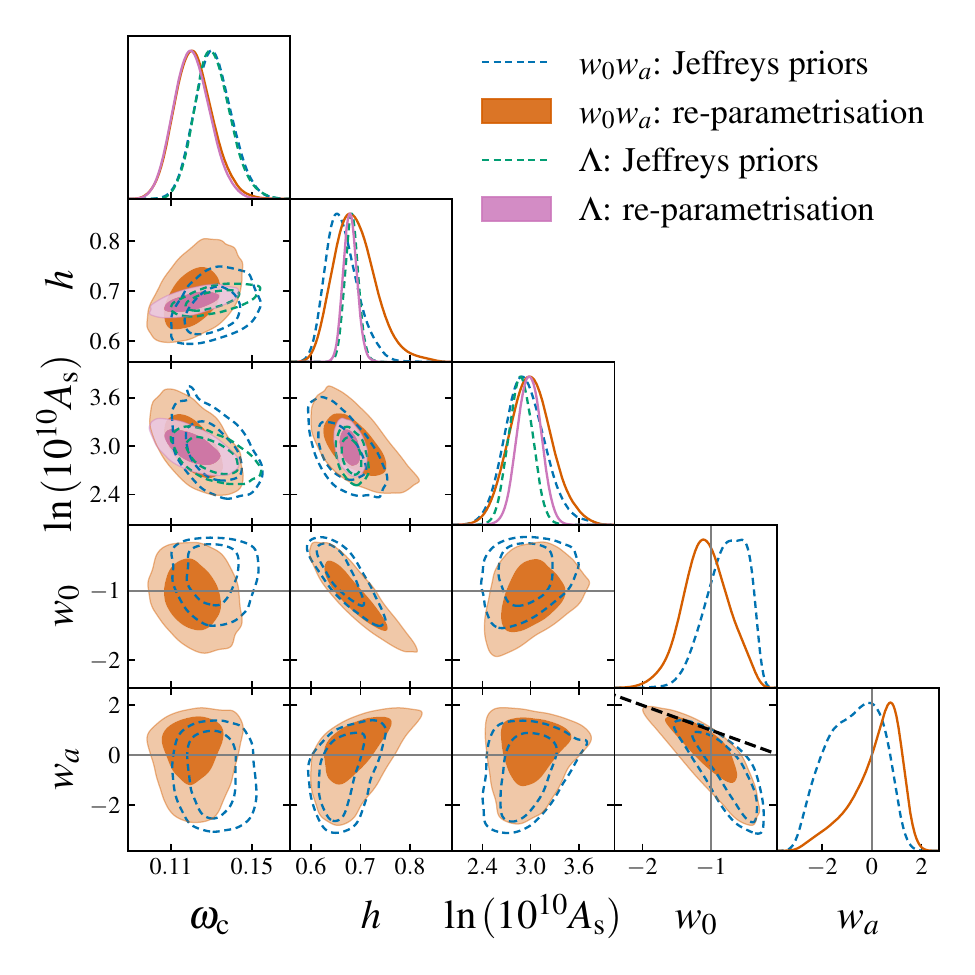}
\caption{BOSS DR12 FS constraints with the re-parametrisation and Jeffreys priors of the analytically marginalised parameters in the standard cosmology and evolving dark energy. Grey solid lines denote the $\Lambda$CDM limit with $w_0=-1$ and $w_a=0$.}
\label{fig:boss-real}
\end{figure}

In appendix~\ref{app:full-reparam}, we investigate versions of a ``full'' re-parametrisation, which includes absorbing power spectrum amplitude parameters into the galaxy bias parameters ($b_1, b_2, b_{\mathcal{G}_2}$) as well. The results are very similar to the simpler re-parametrisation applied only to analytically marginalised parameters.  Hence, we focus on the re-parametrisation of the parameters that appear linearly in the modelling. Another reason for this choice is to directly compare it with the Jeffreys prior approach, that in the cases shown here are only applied to linear parameters. Both approaches are easy to implement and able to mitigate prior-volume effects to a similar degree (especially for Stage IV surveys, as we argue in section~\ref{subsec:stage4}). In contrast to Jeffreys priors, the re-parametrisation is independent from the particular experiment being conducted, since it is fully based on the modelling and does not depend on the covariance. 

\begin{table}[h!]
\footnotesize
  \centering
  \begin{tabular} { l | cc | cc }
    & \multicolumn{2}{c|}{{\bf re-parametrisation}} &  \multicolumn{2}{c}{{\bf Jeffreys priors}}  \\
   & $\Lambda$CDM & $w_0w_a$CDM & $\Lambda$CDM & $w_0w_a$CDM\\
    \hline
	\hline
    $\omega_c$ & $0.121^{+0.008}_{-0.009}$ $(0.124)$ & $0.121^{+0.008}_{-0.010}$ $(0.122)$ & $0.131^{+0.008}_{-0.009}$ $(0.128)$ & $0.131^{+0.008}_{-0.009}$ $(0.126)$\\
    $h$& $0.679^{+0.013}_{-0.014}$ $(0.676)$& $0.690^{+0.034}_{-0.048}$ $(0.673)$ & $0.682\pm 0.014$ $(0.681)$& $0.663^{+0.025}_{-0.037}$ $(0.649)$\\
    $\ln(10^{10}A_s)$ & $2.98\pm 0.15$ $(2.92)$ & $3.00^{+0.25}_{-0.28}$ $(3.25)$ & $2.88\pm 0.15$ $(2.93)$& $2.94^{+0.22}_{-0.27}$ $(3.16)$\\
    $w_0$& $-1$ & $-1.08\pm 0.33$ $(-1.12)$ & $-1$ & $-0.77^{+0.43}_{-0.16}$ $(-0.81)$\\
    $w_a$& $0$ & $0.16^{+1.20}_{-0.55}$ $(0.96)$ & $0$ & $-0.72^{+1.30}_{-0.99}$ $(0.03)$\\
    $\sigma_8$& $0.783\pm 0.050$ $(0.770)$ &$ 0.784\pm 0.053$ $(0.797)$ & $0.785^{+0.048}_{-0.054}$ $(0.798)$ & $0.788^{+0.055}_{-0.064}$ $(0.829)$ \\
    $S_8$& $0.798\pm 0.054$ $(0.798)$ & $0.791^{+0.068}_{-0.077}$ $(0.824)$ & $0.825\pm 0.056$ $(0.832)$ & $0.854\pm 0.081$ $(0.900)$\\
  \end{tabular}

  \caption{Mean values and 68\% c.l. values for $\Lambda$CDM and evolving dark energy with fixed
    neutrino mass $M_{\nu}=0.06~{\rm eV}$ for the FS analysis of BOSS data, using the re-parametrisation and Jeffreys priors on analytically marginalised parameters. We do not show constraints on $n_{\rm s}$ and $\omega_{\rm b}$ as they are completely prior-dominated. We add the maximum values of the un-marginalised posterior in parentheses, and include
    derived constraints on $\sigma_8$ and $S_8$.}
  \label{tab:boss-constraints}
\end{table}

In figure~\ref{fig:boss-real}, we show constraints on the real BOSS DR12 data, with the re-parametrisation and Jeffreys priors on the analytically marginalised parameters. The corresponding values of the posterior average and MAP are provided in table~\ref{tab:boss-constraints}. Analogously to the synthetic data tests, we observe a $1\sigma$-difference in $\omega_{c}$ between the approaches -- both in $\Lambda$CDM and $w_0w_a$CDM. The constraints on the primordial amplitude agree between both approaches in both dark energy scenarios. They are also in better agreement with the Planck values of $\ln{(10^{10}A_{\rm s})}$ \cite{planck2018cosmo} than obtained in the baseline analysis (see figure~\ref{fig:synth-vs-real}). Constraints of the expansion rate are nearly identical in standard cosmology between both approaches. They change when the parameter space is extended to include evolving dark energy parameters. We observe a similar picture as in figure~\ref{fig:boss-anmarg-parameterisation}: $h$-estimate is tighter with Jeffreys priors, $w_a$ is weakly constrained, there is a $0.8\sigma$ difference for marginalised means in $w_0$ between the re-parametrisation and Jeffreys priors. Overall, both approaches agree in inferred constraints on cosmological parameters and significantly reduce projection effects between the expansion rate and evolving dark energy parameters. The reduction in projection effects is also evident from the fact that cosmological parameters corresponding to the maximum of the un-marginalised posterior (values in the parenthesis, obtained with \texttt{minuit}) are close to the maxima of the marginalised posteriors in figure~\ref{fig:boss-anmarg-parameterisation}. This was not not the case in previous analyses with the baseline setup (see refs.~\cite{Carrilho:2022mon, Tsedrik:2025cwc}).  Also note that when the cosmological background changes from $\Lambda$CDM to $w_0w_a$CDM, the central values in cosmological parameters remain unchanged within the errors, both with the re-parametrisation and Jeffreys priors.  Additionally, we see very little evidence for the $\sigma_8$- or $S_8$-tension with respect to CMB measurements (see table~\ref{tab:boss-constraints}): e.g., $S_8=0.832 \pm 0.013$ in $\Lambda$CDM of Planck 2018 \cite{planck2018cosmo}. This consistency between CMB and LSS values of $S_8$ holds true for standard cosmology and evolving dark energy, and agrees with the findings of DESI for DR1 FS \cite{DESI:2024hhd} as well as with the recent results from KiDS-Legacy data \cite{Stolzner:2025htz}.

\subsection{Stage IV surveys}
\label{subsec:stage4}
\begin{figure}[t!]
\centering
\includegraphics[width=0.49\textwidth]{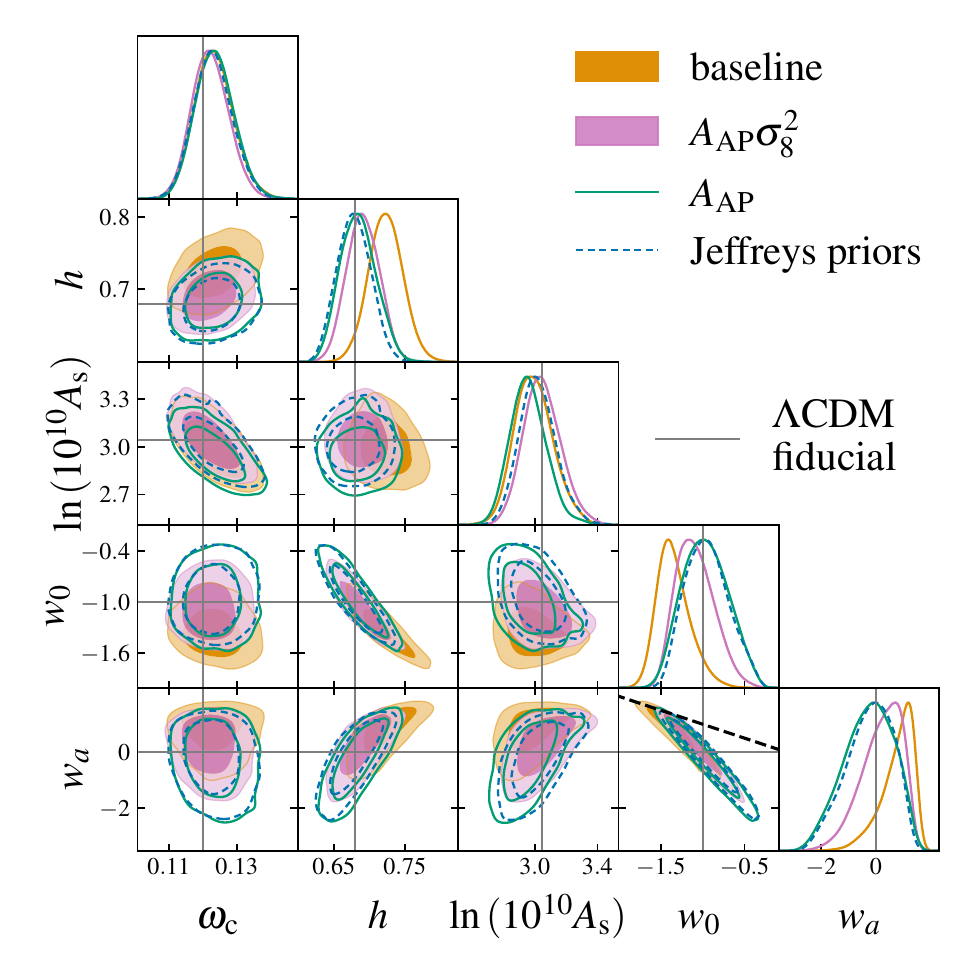}
\includegraphics[width=0.49\textwidth]{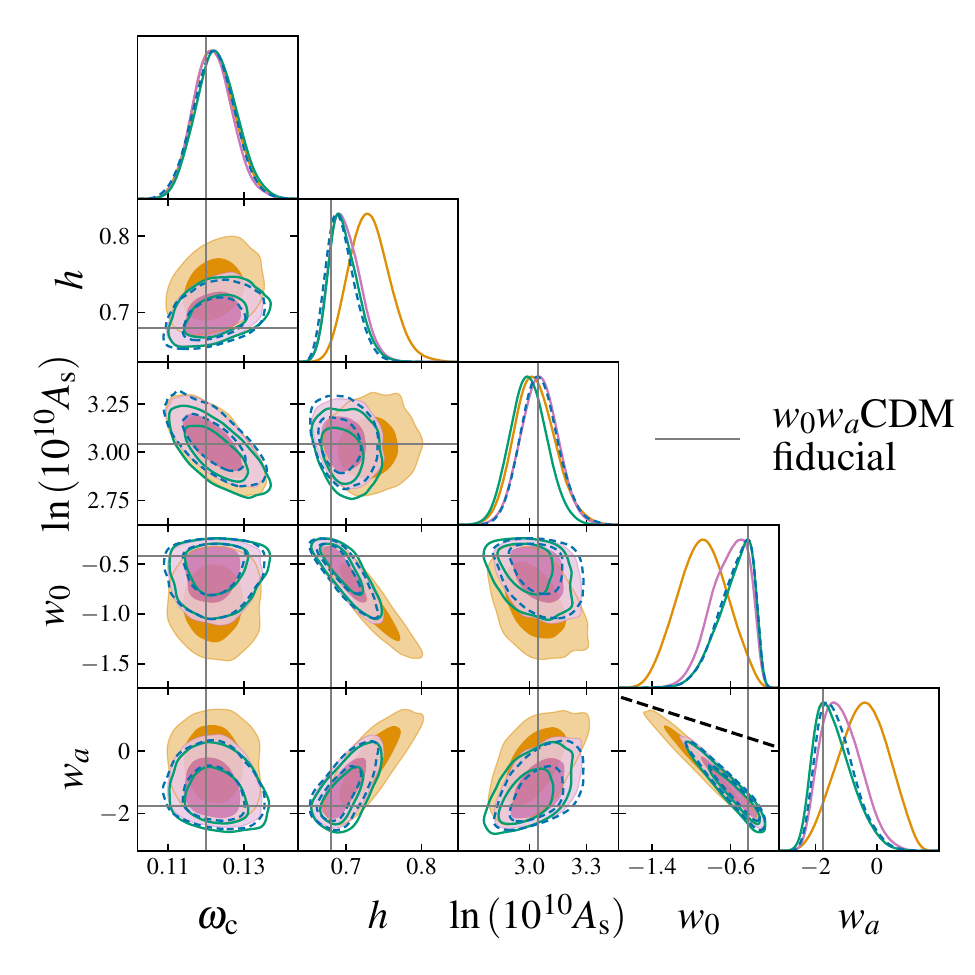}
\caption{Constraining evolving dark energy with synthetic DESI DR1-like data using different FS analysis choices, as specified in the legend (see table~\ref{tab:param-full} and the main text for details on the nuisance parameter priors). \textit{Left panel:} marginalised posterior distribution for cosmological
      parameters with the fiducial $\Lambda$CDM cosmology. \textit{Right panel:} similar to the left panel but with the fiducial $w_0w_a$CDM cosmology. Grey solid lines mark the
      fiducial values of the noiseless synthetic data.}
\label{fig:desi-dr1}
\end{figure}

In this section we repeat the analysis on synthetic data for a DESI DR1-like setup. We re-run our analysis re-scaling the priors on the analytically marginalised parameters as $\mathcal{N}(A_{\rm fid}\mu, A_{\rm fid}\sigma)$ and show results in figure~\ref{fig:desi-dr1}. For the re-scaling with the AP amplitude, we use $A_{\rm fid}=1$, while for the re-scaling with $\sigma_8$, we have $A_{\rm fid} = \sigma_{8,z, \mathrm{fid}}^2$. The latter is slightly less conservative than in the same re-parametrisation for BOSS data, since it decreases the size of the priors on the analytically marginalised parameters taking the re-scaling into account. Of course, this DESI-like setup is an over-simplification of the actual DR1 setup, which includes more redshift bins and a non-Gaussian covariance. However, this proof of concept resembles the official DESI findings \cite{DESI:2024jis} in the baseline analysis. It also shows how robustly our simple re-parametrisation works and effectively re-creates the impact of Jeffreys priors. Additionally, it demonstrates that for measurements with projections less severe than in BOSS for $\omega_c-A_s$, re-scaling with the AP-amplitude might suffice to mitigate $w_0-w_a-h$ projection effects. Finally, to avoid any bias towards the standard cosmology due to the synthetic data-vectors setup, we demonstrate that our re-parametrisation approach works well for a fiducial evolving dark energy scenario too. In this case we set fiducial values for the dark energy parameters to $w_0=-0.42$ and $ w_a=-1.75$, the central values for the deviation from the standard cosmology found in the DESI DR2 BAO+CMB analysis~\cite{DESI:2025zgx}. In the right panel of figure~\ref{fig:desi-dr1}, we again observe an effective mitigation of the projection effects with the re-parametrisation.

\section{BOSS DR12 full-shape and other probes}
\label{sec:external-probes}
\subsection{BAO}
\label{sec:boss-bao}
\begin{figure}[t]
\centering
\includegraphics[width=0.32\textwidth]{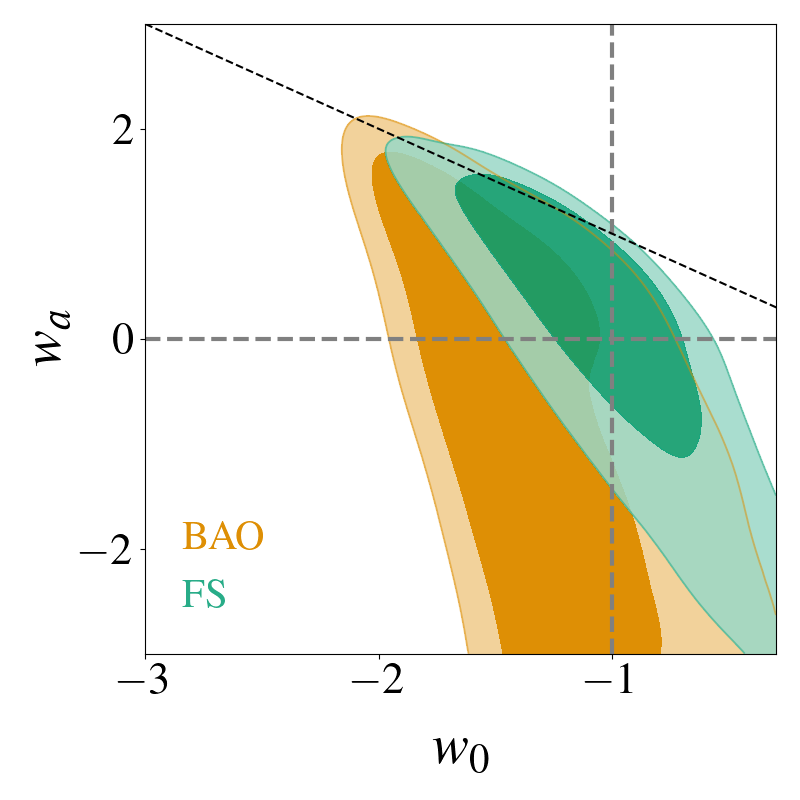}
\includegraphics[width=0.32\textwidth]{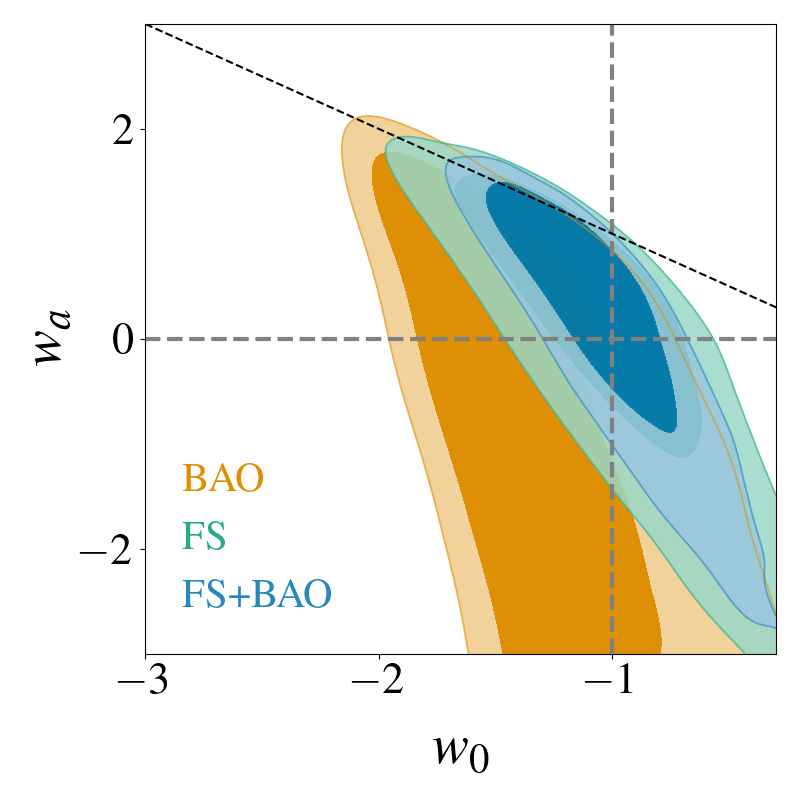}
\caption{Constraining the evolving dark energy parameters with BOSS DR12 probes: BAO and FS alone (left panel) and together with their combination (right panel). Grey dashed lines mark the $\Lambda$CDM limit, black dashed lines denote the condition $w_0+w_a<0$. }
\label{fig:bao-vs-fs}
\end{figure}

After developing and assessing the re-parametrisation approach in the FS analysis, we proceed with constraining evolving dark energy exclusively with pre-DESI LSS probes. First, we contemplate the question whether a FS analysis adds any information on background parameters compared to BAO alone. For this we take four $\alpha$'s from the same sky-cuts as our multipoles and fit background parameters keeping the BBN prior on the baryonic density. It is evident from figure~\ref{fig:bao-vs-fs} that constraints on $w_0-w_a$ are not purely driven by the information in BAO for BOSS, but FS adds constraining power. The FS analysis provides an additional channel to extract information on $\Omega_{\rm m}$, which is relevant for the degeneracies between the evolving dark energy and background evolution parameters. With mitigated projection effects, not only does FS slightly change the degeneracy direction in $w_0-w_a$, but also improves constraints on the time-evolving component $w_a$. This is significant for the BOSS setup, as it contains only two redshift bins $z_1=0.38$ and $z_3=0.61$: i.e. two degrees of freedom in $w(z)$ per two redshifts. It might be interesting to asses the contribution of FS to the BAO constraints of the evolving dark energy parameters in the DESI setup. There, BAO- and FS-measurements span a much broader range of redshifts and the precision of measurements is also significantly improved with respect to BOSS.

\subsection{External large-scale structure probes: BAO and DES Y3}
\label{sec:extbao-des}
\begin{figure}[t]
\centering
\includegraphics[width=0.32\textwidth]{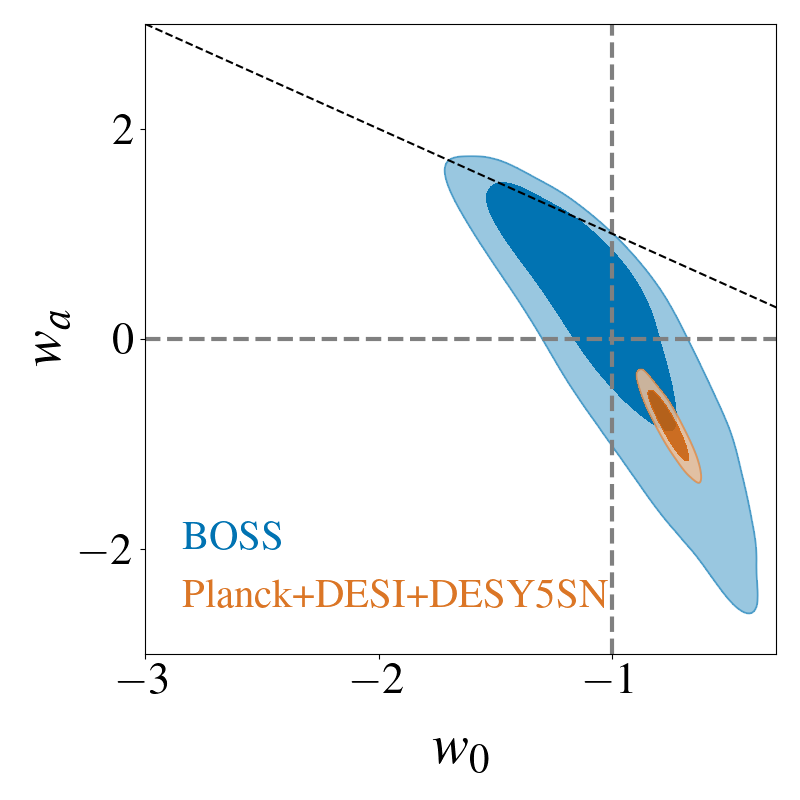}
\includegraphics[width=0.32\textwidth]{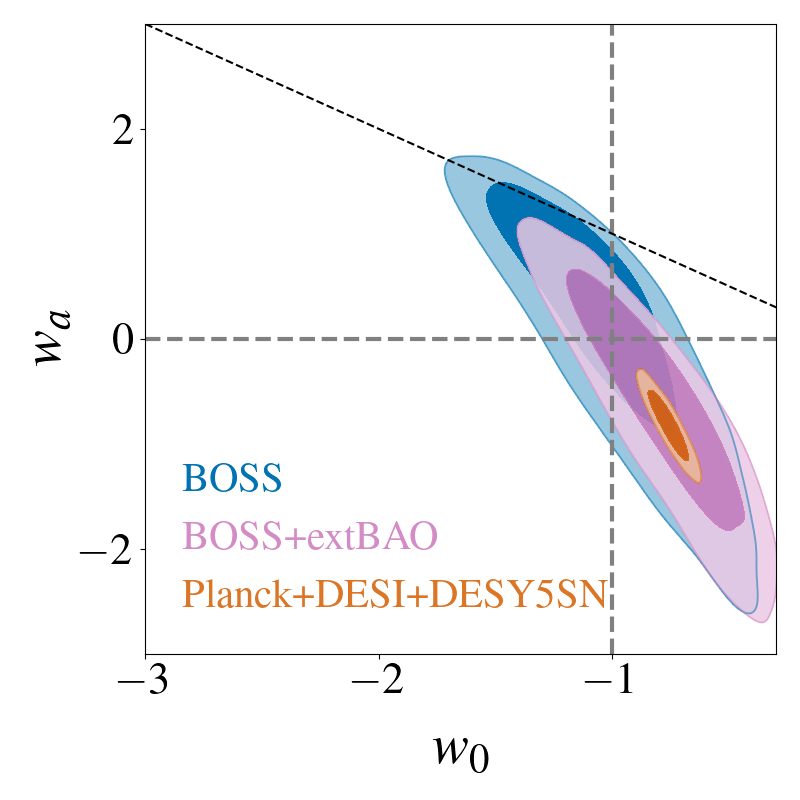}
\includegraphics[width=0.32\textwidth]{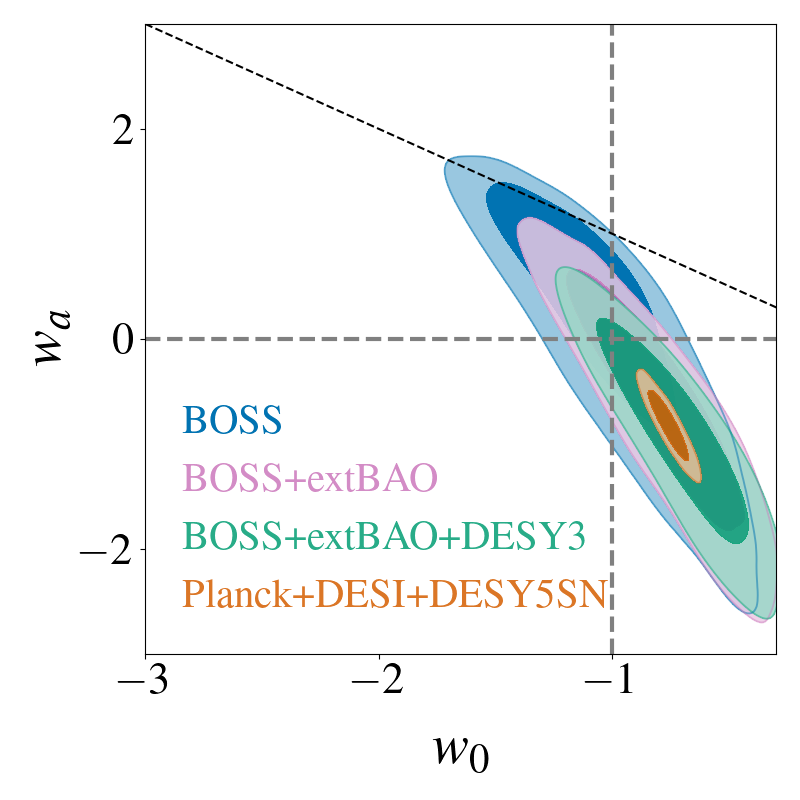}
\caption{Constraining the evolving dark energy parameters with BOSS and external probes: BOSS DR12 FS+BAO (left panel in blue), BOSS and external BAO measurements (middle panel in pink), BOSS and external BAO with the photometric probes from DES Y3 (right panel in green). For comparison, the orange contours presents constraints from DESI DR2 BAO with the CMB and SN from DES Y5. Grey dashed lines mark the $\Lambda$CDM limit, black dashed lines denote the condition $w_0+w_a<0$.}
\label{fig:w0wa}
\end{figure}

After highlighting the value of adding FS information to a BAO-only analysis, we also combine with additional external probes. The resulting constraints on evolving dark energy are shown in figure~\ref{fig:w0wa}, while numerical values are provided in table~\ref{tab:boss-constraints-joint}. The addition of external BAO measurements from lower and higher redshifts than BOSS starts shifting the contours to the parameter space preferred by DESI (central panel of figure~\ref{fig:w0wa}). This happens only if the higher redshift BAO measurements are included in the analysis, providing an anchor in the time evolution. For discussions on the evolving dark energy signatures in DESI see, e.g., refs.~\cite{Cortes:2024lgw, Shlivko:2024llw, DESI:2024aqx,DESI:2024kob, DESI:2025fii, DESI:2025wyn}.

Lastly, we combine Stage III spectroscopic probes with photometric measurements from DES Y3 (left panel of figure~\ref{fig:w0wa}). Note a similar behaviour in the right panel of figure 14 in DESI DR2 BAO analysis~\cite{DESI:2025zgx}: there the authors replace the CMB with the DES Y3 information, to obtain a
constraint coming entirely from low-redshift LSS probes. They also observe that this combination favours the same region of parameter space. For comparison we include contours for a repeated DESI analysis, which in the official collaboration paper constrains evolving dark energy parameters to $w_0=-0.752 \pm 0.057$, $w_a=-0.86^{+0.23}_{-0.20}$ (DESI DR2 BAO+CMB+DESY5) \cite{DESI:2025zgx}. DESI DR1 FS analysis required inclusion of SN, obtaining similar constraints: $w_0=-0.761 \pm 0.065$, $w_a=-1.02^{+0.3}_{-0.26}$ (DESI DR1 FS+BAO+CMB+DESY5) \cite{DESI:2024hhd}. These values remain nearly unchanged if one substitutes DESI DR1 FS with BOSS/eBOSS BAO measurements: $w_0=-0.761 \pm 0.064$, $w_a=-0.88^{+0.29}_{-0.25}$ (DESI DR1 BAO + SDSS BAO + CMB + DESY5) \cite{DESI:2024mwx}. Our purely Stage III LSS driven constraints,
\twoonesig[2cm]{w_0 &= -0.72 \pm 0.21}{w_a &= -0.91^{+0.78}_{-0.64}}{BOSS+extBAO+DESY3, \label{eq:final_w0wa_constraints}}
are in good agreement with DESI results, hinting at a potential deviation from standard cosmology but with larger uncertainties and less significance. 

Moreover, similar to our previous work \cite{Tsedrik:2025cwc}, we advocate for the joint analysis of spectroscopic and photometric probes: FS+BAO+3$\times$2pt. This combination provides a level of constraining power comparable to the addition of CMB information to the spectroscopic analysis. For instance, in ref.~\cite{Lu:2025gki} the authors find constraints similar to ours of $w_0=-0.72^{+0.09}_{-0.11}$ and $w_a=-0.91^{+0.42}_{-0.33}$ for smooth quintessence, i.e. evolving dark energy without clustering. They use a similar set of probes: BOSS DR12 EFTofLSS baseline FS and bispectrum (which presumably mitigates projections) at one-loop level, with external BAO (nearly the same as ours, the high redshift ones are eBOSS DR14 instead of DR16), and Planck instead of DES Y3. Last but not least, a comparison to the official BOSS DR12 results: with the inclusion of CMB information from Planck, FS+BAO (with no extBAO) constrain $w_0=-0.68\pm 0.18$ and $w_a=-0.98 \pm0.53$ \cite{BOSS:2016wmc}. Note that FS in this case is not the EFTofLSS approach of fitting the multipoles, but a compression of the multipoles and clustering wedges statistics, in configuration and Fourier space, into the following properties \cite{BOSS:2016teh, BOSS:2016hvq, BOSS:2016off, BOSS:2016ntk}: $D_{\rm M}(z)/r_{\rm d}$, $H(z)r_{\rm d}$, $f(z) \sigma_8(z)$. These different approaches are later combined in a consensus set and then used in the analysis. Additionally, in this analysis no projection effects were observed, as the different methodologies applied to BOSS data resulted in posteriors that were well described by Gaussian multivariate distributions. The redshift binning is also different: it includes an additional intermediate $z$-bin, $z_2=0.51$. Agreement with these Stage III constraints, while not manifesting a significant deviation from $\Lambda$CDM, showcases a successful application of our re-parametrisation in FS analysis.

\begin{table}[h!]
\footnotesize
  \centering
  \begin{tabular} { l | c | c | c | c}
     &  BOSS BAO & BOSS FS+BAO & BOSS+extBAO & BOSS+extBAO+DES\\
    \hline
	\hline
    $\omega_c$ & $0.249^{+0.073}_{-0.053}$ $(0.280)$ & $0.121^{+0.008}_{-0.010}$ $(0.122)$& $0.116^{+0.007}_{-0.009}$ $(0.117)$ & $0.119^{+0.005}_{-0.006}$ $(0.120)$\\
    $h$& $0.800^{+0.071}_{-0.026}$ $(0.87)$ & $0.688^{+0.030}_{-0.035}$ $(0.673)$& $0.660^{+0.021}_{-0.028}$ $(0.648)$ & $0.659^{+0.019}_{-0.023}$ $(0.647)$\\
    $\ln(10^{10}A_s)$ & - & $3.03^{+0.20}_{-0.24}$ $(3.05)$& $3.05\pm 0.18$ $(3.02)$ & $2.99\pm 0.10$ $(2.99)$ \\
    $w_0$& $-1.34^{+0.33}_{-0.47}$ $(-1.9)$ & $-1.06^{+0.23}_{-0.31}$ $(-1.02)$& $-0.80\pm 0.24$ $(-0.68)$& $-0.72\pm 0.21$ $(-0.64)$\\
    $w_a$& $< 0.169$ $(-0.8)$ & $0.22^{+1.10}_{-0.46}$ $(0.44)$& $-0.63^{+0.92}_{-0.72}$ $(-0.90)$ & $-0.91^{+0.78}_{-0.64}$ $(-1.03)$\\
    $\sigma_8$& - & $0.786\pm 0.049$ $(0.783)$& $0.777\pm 0.048$ $(0.768)$ & $0.773\pm 0.023$ $(0.765)$ \\
    $S_8$& - & $0.793\pm 0.058$ $(0.809)$& $0.801\pm 0.055$ $(0.810)$ & $0.809\pm 0.016$ $(0.818)$ \\
  \end{tabular}
  \caption{Mean values and 68\% c.l. values for the $w_0w_a$CDM cosmology with fixed
    neutrino mass $M_{\nu}=0.06~{\rm eV}$ for BOSS BAO (with the upper bound on $w_a$ referring to the 68\% limit), FS+BAO with BOSS, their combination with external BAO measurements, as well as with DES Y3 3$\times$2pt analysis. We show the best-fit values in parentheses, and include
    derived constraints on $\sigma_8$ and $S_8$.}
  \label{tab:boss-constraints-joint}
\end{table}

\section{Conclusion}
\label{sec:conclusion}

The goal of this paper is two-fold: 1) to present a simple re-parametrisation approach for the EFTofLSS full-shape analysis to mitigate projection effects, and 2) to apply this method in combination with other external probes from Stage III large-scale low-redshift measurements (pre-DESI). First, we pedagogically introduced projection or prior-volume effects. These effects manifest themselves through shifts in the marginalised posterior maxima from the corresponding best-fit values of the un-marginalised posterior. In general, they affect any non-Gaussian multi-dimensional posterior distribution that is compressed into a lower dimensional parameter space. In case of the full-shape approach this comes from the degeneracy between cosmological, extended (beyond-$\Lambda$CDM) and the EFTofLSS nuisance parameters -- all of them affecting the amplitude of the measured power spectrum multipoles. We introduce a re-parameterisation of the EFTofLSS nuisance parameters with the amplitudes $A_{\rm AP}\sigma_{8, z}^2$ (the Alcock-Paczynski amplitude times the variance of the density field, smoothed within $8$ Mpc $h^{-1}$ and re-scaled with the time-dependent growth factor), and show it weakens the impact of projection effects. We tested this approach on synthetic noiseless data generated for BOSS DR12 and DESI DR1 setups in the evolving dark energy scenario, $w_0w_a$CDM. For both setups, we compared the performance of the re-parametrisation to an alternative method of mitigating projection effects -- Jeffreys priors. Imposing such priors and re-parametrising the same set of the nuisance parameters, both approaches showed a good level of agreement. The advantages of the re-parametrisation approach: it is simple to implement and interpret, straightforward to extend to nuisance parameters which appear nonlinearly in the model, and is independent from the particular experiment because it is fully based on the modelling and does not depend on the covariance. Beyond the purely frequentist approach (for instance, see the recent ref.~\cite{DESI:2025hao}), the re-parametrisation is one of the simplest ways to measure evolving dark energy with mitigated prior-volume effects in the Bayesian framework.  

In the second half of the paper, we focused on constraining evolving dark energy with Stage III publicly available data: BOSS DR12 (power spectrum multipoles and BAO), external BAO measurements (from 6DF, SDSS DR7 MGC and eBOSS DR16 surveys), and the 3$\times$2pt correlation functions of DES Y3. We found that, for BOSS data, the full-shape analysis with the re-parametrisation adds information on dark energy parameters with respect to the BAO-only analysis. We also found that adding BAO information from redshifts higher than those covered by BOSS drives the constraints into the parameter region preferred by DESI, namely $w_0>-1$ and $w_a<0$. Finally, when combining all probes together under the application of the re-parametrisation in the full-shape analysis, we obtained the following constraints of the evolving dark energy parameters: $w_0=-0.72 \pm 0.21$ and $w_a=-0.91^{+0.78}_{-0.64}$. To our knowledge, these are the first purely large-scale structure pre-DESI constraints in the $w_0w_a$CDM cosmology. They are also in good agreement with the official BOSS DR12 results \cite{BOSS:2016wmc} and a recent analysis with the one-loop bispectrum \cite{Lu:2025gki} -- both analyses include CMB information from Planck instead of the 3$\times$2pt correlation functions of DES Y3. Although our constraints do not indicate a significant deviation from $\Lambda$CDM, they still demonstrate robustness of our re-parametrisation and potential of exclusively late-time Universe constraints.

The methods and findings of this study are important for forthcoming beyond-$\Lambda$CDM analyses by Stage IV surveys like DESI, Euclid, Rubin, and their combinations.

\appendix

\section{Full re-parametrisation}
\label{app:full-reparam}
\begin{table}[h!]
\footnotesize
  \centering
  \begin{tabular} {  c c | c c}
   \multicolumn{2}{c|}{{\bf ``Kaiser term'' re-parametrisation}} & \multicolumn{2}{c}{{\bf ``ignore Kaiser term'' re-parametrisation}} \\ 
    full+Kaiser &  full+Kaiser+AP & full  &  full+AP \\
    \hline
	\hline
        $b_1 \sigma_{8, z}$ &  $b_1 \sigma_{8, z} \sqrt{A_{\rm AP}}$ &  $b_1 \sigma_{8, z}$ &  $b_1 \sigma_{8, z} \sqrt{A_{\rm AP}}$ \\
    $b_2 \sigma_{8,z}^2$ & $b_2 \sigma_{8,z}^2 \sqrt{A_{\rm AP}}$ &  $b_2 \sigma_{8,z}^2$ & $b_2 \sigma_{8,z}^2 \sqrt{A_{\rm AP}}$ \\
    $b_{\mathcal{G}_2}\sigma_{8,z}^2$ & $b_{\mathcal{G}_2}\sigma_{8,z}^2 \sqrt{A_{\rm AP}}$ &  $b_{\mathcal{G}_2}\sigma_{8,z}^2$ & $b_{\mathcal{G}_2}\sigma_{8,z}^2 \sqrt{A_{\rm AP}}$ \\
    \hline
      $b_{\Gamma_3} A_{\rm AP} \sigma_{8, z}^3 $ & $b_{\Gamma_3} \sqrt{A_{\rm AP}} \sigma_{8, z}^3 $ &  $b_{\Gamma_3} A_{\rm AP} \sigma_{8, z}^4 $ & $b_{\Gamma_3} A_{\rm AP} \sigma_{8, z}^4 $ \\
      $c_0 A_{\rm AP}\sigma_{8, z}^2$ & $c_0 A_{\rm AP}\sigma_{8, z}^2$ &  $c_0 A_{\rm AP}\sigma_{8, z}^2$ & $c_0 A_{\rm AP}\sigma_{8, z}^2$ \\
    $c_2 A_{\rm AP}\sigma_{8, z}^2$ & $c_2 A_{\rm AP}\sigma_{8, z}^2$ & $c_2 A_{\rm AP}\sigma_{8, z}^2$ & $c_2 A_{\rm AP}\sigma_{8, z}^2$ \\
    $c_4 A_{\rm AP}\sigma_{8, z}^2$ & $c_4 A_{\rm AP}\sigma_{8, z}^2$ &  $c_4 A_{\rm AP}\sigma_{8, z}^2$ & $c_4 A_{\rm AP}\sigma_{8, z}^2$ \\
    $c_{\nabla^4\delta} A_{\rm AP}$ & $c_{\nabla^4\delta}$ &  $c_{\nabla^4\delta} A_{\rm AP}\sigma_{8, z}^2$ & $c_{\nabla^4\delta} A_{\rm AP}\sigma_{8, z}^2$ \\
    $NA_{\rm AP}$ & $NA_{\rm AP}$&$NA_{\rm AP}$ & $NA_{\rm AP}$\\
    $e_0 A_{\rm AP}$ & $e_0 A_{\rm AP}$  & $e_0 A_{\rm AP}$ & $e_0 A_{\rm AP}$\\
    $e_2 A_{\rm AP}$ & $e_2 A_{\rm AP}$ &$e_2 A_{\rm AP}$ & $e_2 A_{\rm AP}$ 
  \end{tabular}

  \caption{Various options to re-parametrise the EFTofLSS nuisance parameters, both analytically marginalised and sampled explicately (separated by the horizontal line).}
  \label{tab:param-full}
\end{table}
\begin{figure}[t]
\centering
\includegraphics[width=0.55\textwidth]{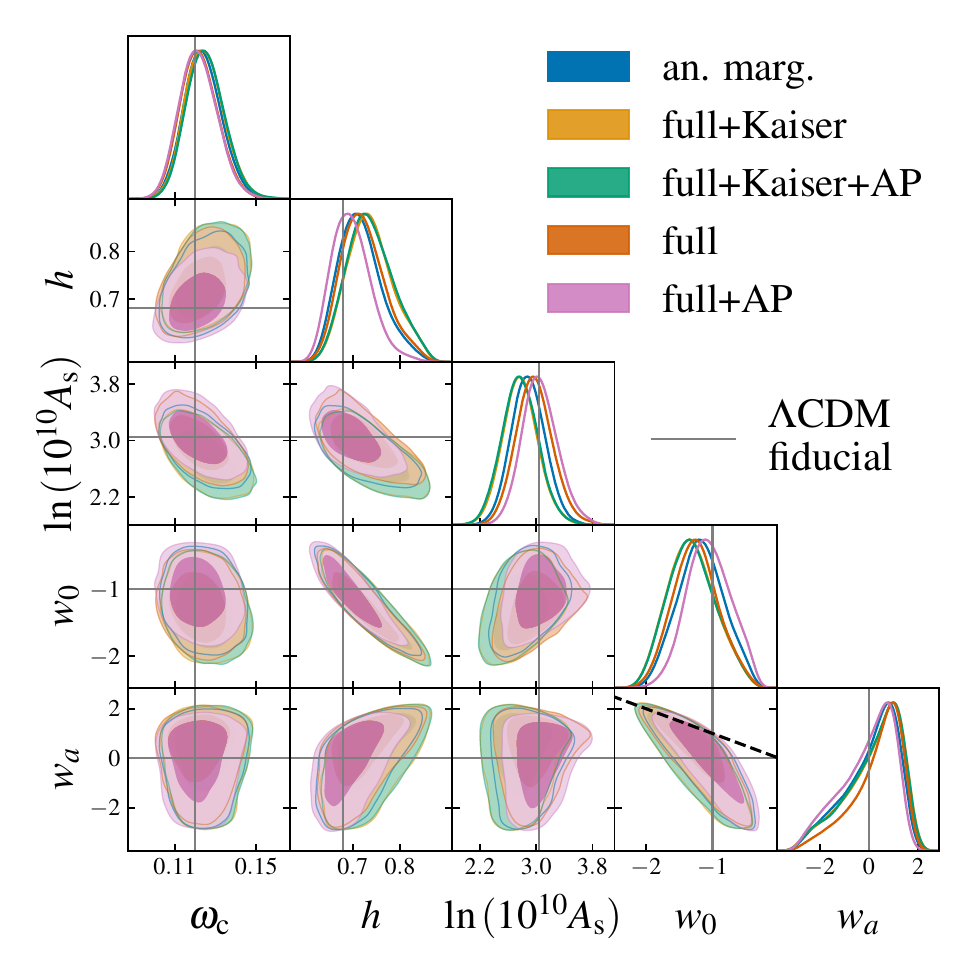}
\includegraphics[width=0.3\textwidth]{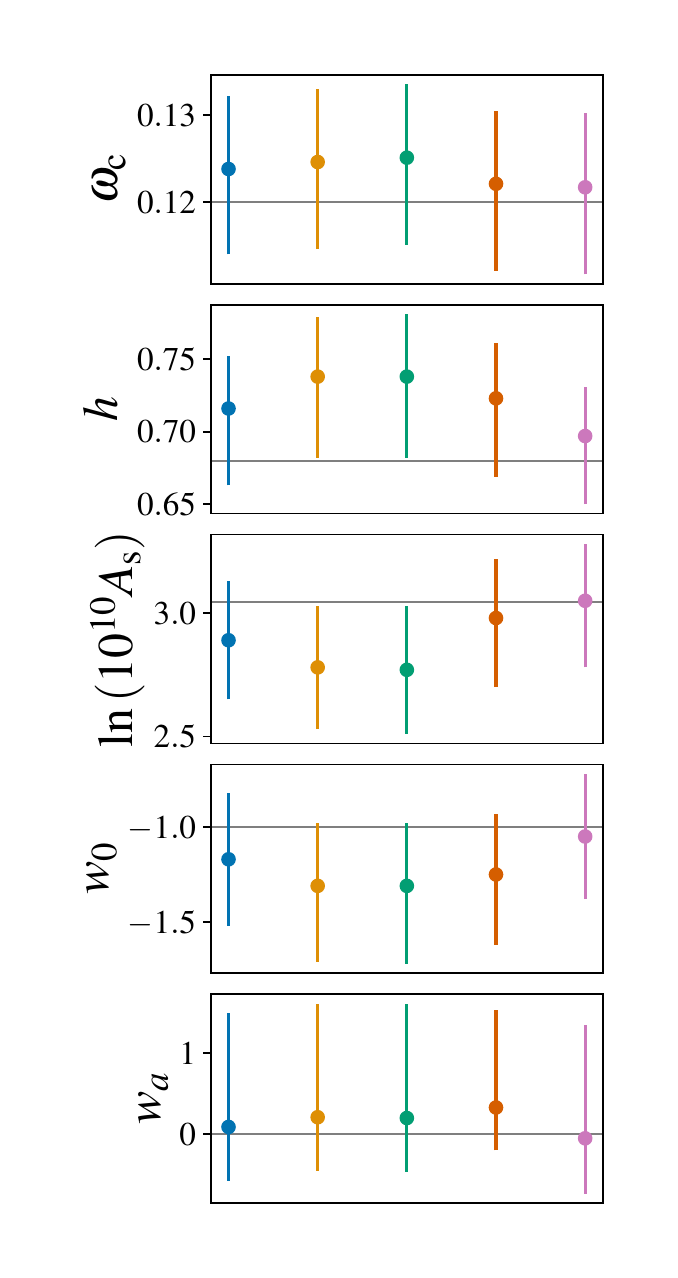}
\caption{Different re-parametrisations of the EFTofLSS nuisance parameters and their impact on cosmological constraints demonstrated on synthetic data. \textit{Left panel:} marginalised posterior distribution for the cosmological
      parameters in the $w_0w_a$CDM cosmology and the five re-parametrisation choices,
      as detailed in the legend (blue colour denotes the case of $A_{\rm AP}\sigma^2_{8,z}$ re-scaling of the analytically marginalised parameters from table~\ref{tab:param-priors}, other colours denote the options from table~\ref{tab:param-full}). Grey solid lines mark the
      fiducial values of the noiseless synthetic data from a fit on the BOSS DR12 multipoles in a $\Lambda$CDM scenario. \textit{Right panel:} errorbars correspond to 68\% c.l. of 1D marginalised constraints on cosmological parameters from the left panel.}
\label{fig:boss-full-parameterisation}
\end{figure}

In a full re-parametrisation of the EFTofLSS nuisance parameters, not analytically marginalised galaxy bias parameters also re-absorb an amplitude parameter. Similar to DESI's approach, we obtain $b_1 \rightarrow b_1 \sigma_{8,z}$ and $b_{2, \mathcal{G}_2} \rightarrow b_{2, \mathcal{G}_2} \sigma_{8,z}^2$. To take the AP-amplitude into account (e.g., in term like $b_1^2A_{\rm AP}P_{\rm L}$) we add a square-root of it to the newly defined galaxy bias parameters. Note that we have various powers of these galaxy bias parameters in the model. Hence, the cancellation of the amplitude-controlling parameters will not be as straight-forward as in the case of the analytically marginalised parameters, which appear only linearly in the EFTofLSS. We also note that in the analytically marginalised part we have certain terms multiplied with the Kaiser term, i.e. $b_{\Gamma_3}(b_1+ f\mu^2) P_{13}$ and $c_{\nabla^4\delta}(b_1+f\mu^2)^2P_{\rm L}$. Therefore, we distinguish between two full re-parametrisation approaches: a) one where the Kaiser term is taken into account, i.e. only the leading order terms, $b_{\Gamma_3}b_1 A_{\rm AP}P^2_{\rm L}$ and $c_{\nabla^4\delta}b_1^2 A_{\rm AP}P_{\rm L}$, are multiplied with the amplitude; b) one where we use the same re-parametrisation of the analytically marginalised parameters as before in table~\ref{tab:param-priors}. All options are summarised in table~\ref{tab:param-full}. The results from synthetic BOSS data are shown in figure~\ref{fig:boss-full-parameterisation}. The priors applied in the analyses for each option in figure~\ref{fig:boss-full-parameterisation} are identical to the $A_{\rm AP} \sigma^2_{8,z}$ re-parametrisation analysis from tab.~\ref{tab:param-priors}. We see that all approaches are consistent between each other. The best agreement with the fiducial cosmology is reached by the full re-parametrisation in which the AP-amplitude is absorbed in all nuisance parameters and the analytically marginalised parameters are treated in the same way as presented in table~\ref{tab:param-priors}. Additionally, in the next section we demonstrate stability of this re-parametrisation to changes of the prior-volume.

\section{Prior-dependence}
\label{app:priors}

\begin{figure}[h!]
\centering
\includegraphics[width=0.49\textwidth]{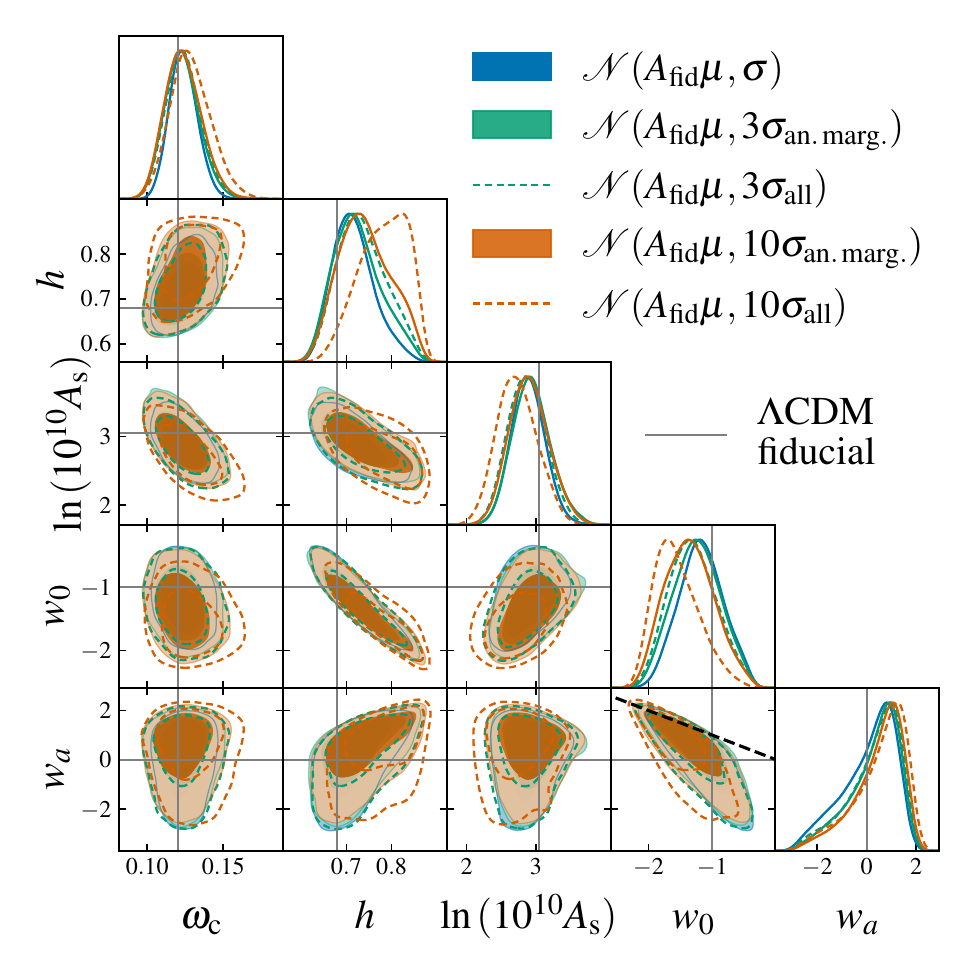}
\includegraphics[width=0.49\textwidth]{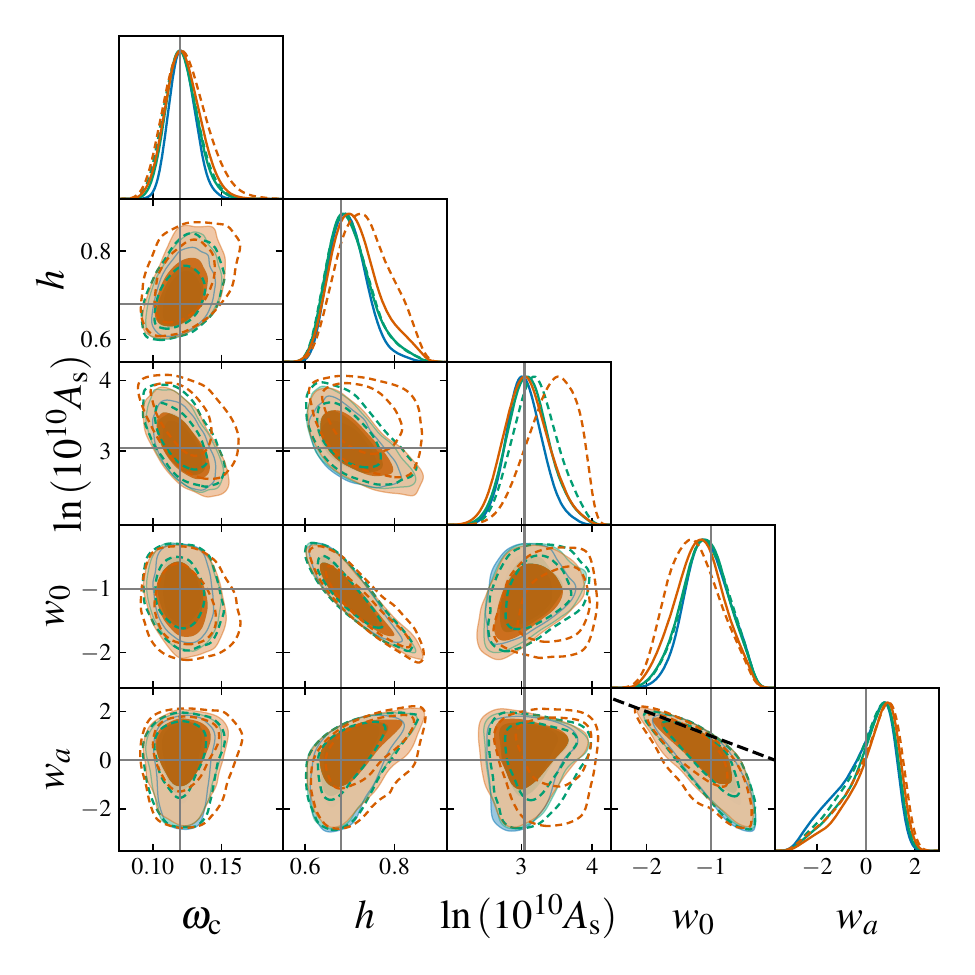}
\caption{Different size of priors of the EFTofLSS nuisance parameters and their impact on cosmological constraints demonstrated on synthetic data. \textit{Left panel:} re-parametrisation of the analytically marginalised parameters. \textit{Right panel:} full re-parametrisation (``full+AP'' in figure~\ref{fig:boss-full-parameterisation}), same colour scheme as in the left panel. \textit{Legend:} the short-hand notation for varying size of the Gaussian priors includes $\mathcal{N}$ with the average $\mu$ and $\sigma$ from the baseline analysis, $A_{\rm fid}$ is the re-scaling factor of the prior-means in the re-parametrisation with $A_{\rm AP}\sigma^2_{8,z}$ (see table~\ref{tab:param-priors}), $\sigma_{\rm an.\, marg.}$ denotes increased priors only in the analytically marginalised parameters, while $\sigma_{\rm all}$ denotes increased priors in all nuisance parameters.}
\label{fig:boss-synth-priors}
\end{figure}

In appendix of ref.~\cite{Carrilho:2022mon}, the authors demonstrate a strong dependence of cosmological constraints on the size of the EFTofLSS priors. We repeatedly vary the size of the Gaussian priors on the nuisance parameters in our re-parametrisation on the $\Lambda$CDM synthetic BOSS-data here. The only difference with respect to ref.~\cite{Carrilho:2022mon}, is imposing a $10\sigma_{\rm Planck}$ prior on $n_{\rm s}$. In figure~\ref{fig:boss-synth-priors} we demonstrate that the re-parametrisation with $A_{\rm AP}\sigma^2_{8,z}$ is nearly prior-independent. One can see only slight shifts when priors on the analytically marginalised parameters are increased. The only case of notable shifts is when priors on $b_{\mathcal{G}_2}$ and $b_2$ are increased 10 times: in $w_0-h$ for the re-parametrisation of the analytically marginalised parameters and in $h-\ln{(10^{10}A_{\rm s})}$ for the full re-parametrisation. Both these scenarios correspond to fairly unrealistic values for perturbative galaxy biases.

\acknowledgments
We thank Alkistis Pourtsidou for useful discussions and an MCMC chain with the DESI DR2 BAO re-analysis. We thank Joe Zuntz for valuable comments. We also thank Samuel Brieden for recommendations on the DESI DR1-like setup. The authors are grateful to the BOSS and DES collaborations for making their data publicly available. We acknowledge use of the Cuillin computing cluster
of the Royal Observatory, University of Edinburgh. MT's research is supported by grant ST/Y000986/1. PC's research was supported by grant RF/ERE/221061 in the initial stages of this work.
For the purpose of open access, the author
has applied a Creative Commons Attribution (CC BY) licence to any
Author Accepted Manuscript version arising from this submission.




\bibliographystyle{JHEP}
\bibliography{biblio.bib}

\end{document}